\documentclass[preprint,showpacs,preprintnumbers,amsmath,amssymb]{revtex4}

\usepackage{graphicx}
\usepackage{dcolumn}
\usepackage{bm}
\usepackage{color}

\usepackage{feynmf}
\usepackage{slashed}
\usepackage{enumerate}

\usepackage[utf8]{inputenc}

\begin{document}

\title{Calculations of the Sommerfeld Effect in a Unified Wave Function Framework}


\author{Yi-Lei Tang}
\thanks{tangyilei@kias.re.kr}
\affiliation{School of Physics, KIAS, 85 Hoegiro, Seoul 02455, Republic of Korea}

\author{Gao-Liang Zhou}
\thanks{zhougl@itp.ac.cn}
\affiliation{Xi'an University of Science and Technology, 58 Yanta Road, Xi'an, Shaanxi Province, People's Republic of China}

\date{\today}

\begin{abstract}
In this paper, we suggest a method of a complete calculation of the Sommerfeld effect in both the s-wave and p-wave annihilation situations. The basic idea is to uniformly consider the equal-time Beth-Salpeter wave functions of both the dark matter and the final state particles, then short-distance interactions can be parametrized as a cross-term considering the equations. Solving these equations will result in the complete formulas of the cross sections with the Sommerfeld effects.

\end{abstract}
\pacs{}

\keywords{dark matter, relic abundance, sterile neutrino}

\maketitle
\section{Introduction}

If the dark matter particles interact with themselves through exchanging some bosonic light mediator, the annihilation behaviour of a dark matter particle pair as a function of their relative velocity can be significantly altered. If the force that the mediator carries is attractive, then the annihilation cross section $\langle \sigma v \rangle$ will be amplified. This is usually called the Sommerfeld enhancement in the literature (For the original work by A. Sommerfeld, see \cite{SommerfeldOriginal}. For some early applications in the dark matter, see \cite{Hisano0, Hisano1, Hisano2, Hisano3, EarlySommer1, EarlySommer2, Minimal1, Minimal2, EarlySommer3, EarlySommer4, EarlySommer5, EarlySommer6, EarlySommer7, NimaSommerfeld}). Usually, it is the resonance effect of the ``zero-energy bound state''.
 
\begin{figure}
\includegraphics[width=0.65\textwidth]{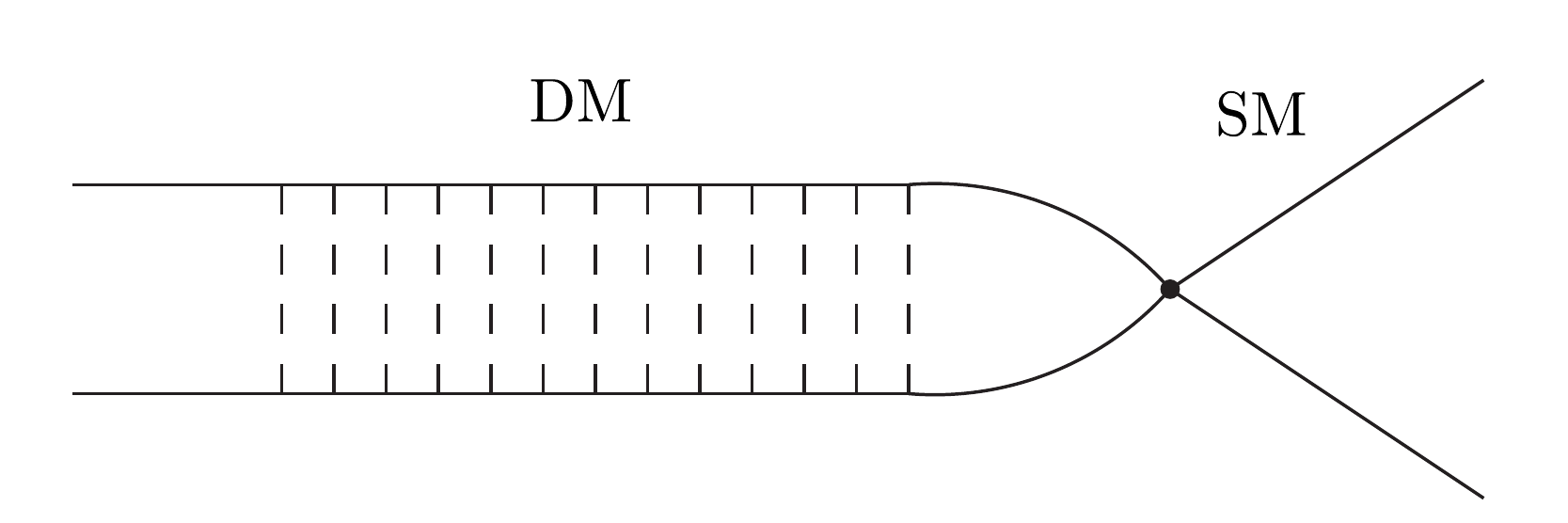}
\caption{Dark matter exchanging light force mediators and finally annihilate into two SM particles. \label{Traditional}}
\includegraphics[width=\textwidth]{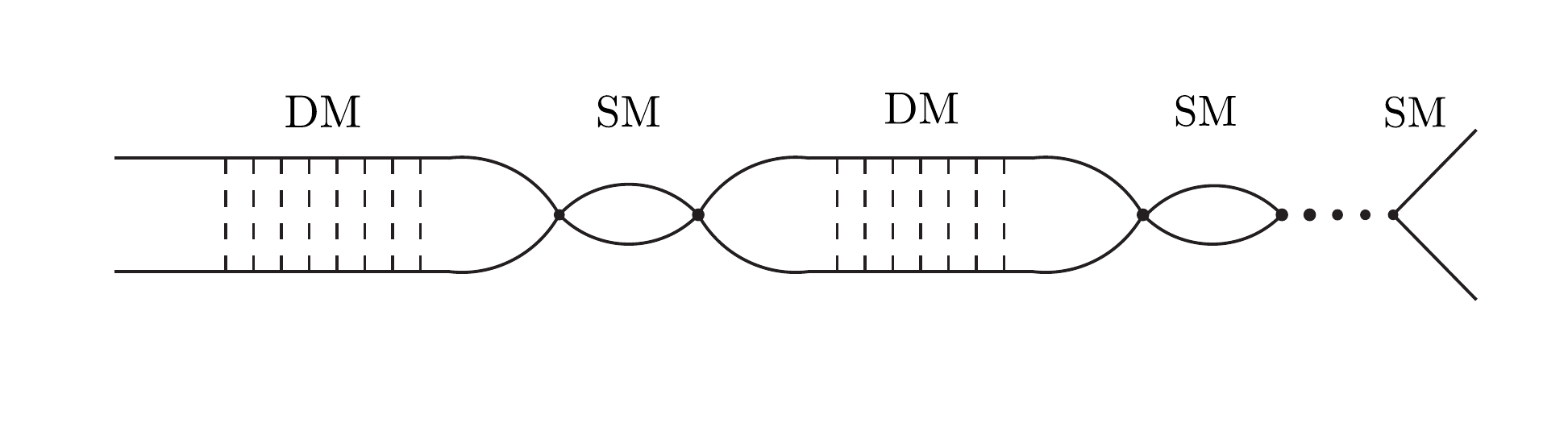}
\caption{An example of the diagrams in which the dark matter particles annihilate to the SM particles, re-annihilate to the dark matter particles, to the SM particles alternatively. \label{Resummation}}
\end{figure}

Traditional calculations of the Sommerfeld effects usually involve solving the Schrödinger equation only including the ``long-distance effects'' at first, and then calculating the perturbative ``short-distance'' cross section $\sigma v_0$. Usually, in the literature, people decompose $\sigma v_0 = a + b_s v^2 + b_p v^2$ in the non-relativistic limit, and $a + b_s v^2$ indicate the s-wave contributions, and $b_p v^2$ is the p-wave contributions. Notice that in the s-wave annihilation, a $v^2$ dependence can still exist. Calculate the the ``boost factor'' $S_s(v)$ and $S_p(v)$, The final result of the cross section becomes $\sigma v = S_s(v) (a + b_s v^2) + S_p(v) b_p v^2$. Here $S_s(v)$ is calculated to be $|\psi(\vec{0})|^2$, where the $\psi(\vec{0})$ is the zero-point wave function of a stationary scattering state, and $S_p(v) = \frac{|\vec{\nabla} \psi(\vec{0})|^2}{k^2}$, where $k$ is the relative momentum of the initial states. This is equivalent to calculating the Feynmann diagrams as shown in the Fig.~\ref{Traditional}. In this paper, we try to regard both these interactions in a unified framework of wave functions in which both the Sommerfeld effect and the annihilation processes are all calculated by solving the equations of wave function. This is equivalent to resumming all the ``trail''-diagrams as in the Fig.~\ref{Resummation}. In the literature, the Ref.~\cite{MainDependence} had followed this idea by introducing a complex $\delta$-function in the dark matter potential (Ref.~\cite{Hisano1, ZREFT} have also adopted this potential), indicating a short-distance coupling making the Hamiltonian non-Hermitian. The $\delta$-function in the quantum mechanics requires renormalization\cite{Jackiw}. The Ref.~\cite{MainDependence} had also pointed out that the traditional calculation method can break the unitarity in some particular parameter space\cite{Unitarity}, and their calculations can cure this problem. This technique is effective in calculating the s-wave annihilation processes.

The p-wave annihilation for the indirect detection signals (sometimes with the Sommerfeld enhancement) have also been discussed in the literature\cite{pWave1, pWave2, pWave3, NewDas}. Generalizing the method in the Ref.~\cite{MainDependence} to the p-wave case confronts with difficulties. $\delta$ function potential only interact with the $l=0$ partial waves. If we simply add some complex potential, e.g., $\frac{ \partial \delta^{(3)} (\vec{r})}{\partial z}$, that interacts with the $l=1$ partial wave, since usually $\langle l=1 | \frac{ \partial \delta^{(3)} (\vec{r})}{\partial z} | l=0 \rangle \neq 0$, such a term will disturb the $l=0$ partial waves. This does not happen in some practical dark matter model.

In this paper, we will make another approach. If we uniformly describe the annihilating dark matter and the final state particles in the form of Bethe-Salpeter wave functions, we can then parametrize and write down the short-distance interaction terms between them in the wave function equations. In this situation, an interaction term like $\frac{ \partial \delta^{(3)} (\vec{r})}{\partial z}$ couples with the different particle states, thus avoiding the problems described before.
 
The Sommerfeld effect arises in the non-relativistic limit of the dark matter, so the dark matter can be described by the simple Schrödinger equation. Usually, the annihilation products are usually light standard model (SM) particles compared with the dark matter particles, and can be regarded as massless particles. It is convenient to describe the final-state particles with the massless Klein-Gordan equations, or d`Alembert equation with an interaction term. We will prove in our paper that the Beth-Salpeter wave function of the relative motion of a pair of particles in many cases satisfy the d’Alembert equation. We will solve these equations and extract the cross sections from the results. We will show the final expressions as well as the derivation processes in detail.

In section \ref{GeneralDiscussion}, we prepare the mathematical tools for the further calculations. The s-wave scattering process is calculated in section \ref{SWaveCalc}, and the p-wave scattering process is calculated in section \ref{PWaveCalc}. In the sub-section \ref{Discussions}, we briefly compare our wave function results with the diagramatic analysis. In the section \ref{WaveFunction}, we prove the validity of the wave function equations we rely on. Finally, we summarize our paper by section \ref{Summary}.

\section{Some General Discussions} \label{GeneralDiscussion}

Let $\chi(\vec{r})$ be the dark matter's stationary wave function of the relative motion, and $\phi(\vec{r})$ be the stationary wave function of relative motion for the final state, where $\vec{r}$ is the relative displacement between the particles inside the pair. The short-distance scattering effects can be parametrized by an interaction term which is non-zero only at the origin. For an s-wave scattering process, only the $\delta$-function arises. Considering the long-distance effects by exchanging some light mediators between the dark matter particles, the stationary equations of the dark matter-annihilation product system can be given by
\begin{eqnarray}
-\frac{\vec{\nabla}^2}{2 \mu} \chi(\vec{r})+V(\vec{r}) \chi(\vec{r}) + u_s \delta^{(3)} (\vec{r}) \phi(\vec{0}) &=& E_k \chi(\vec{r}) \nonumber \\
-\vec{\nabla}^2 \phi(\vec{r}) + \mu u_s \delta^{(3)}(\vec{r}) \chi(\vec{0}) &=& \frac{E_t^2}{4} \phi(\vec{r}), \label{sFunction}
\end{eqnarray}
for s-wave interactions. Note that the second equation in (\ref{sFunction}) is the d'Alembert equation with an interaction term. This is unusual in the literature, and we will leave the proof of it in section \ref{WaveFunction}.

For a p-wave process, the derivative of a $\delta$-function is utilized. In a practical model, the index of the derivative should be contracted with another index. In this paper, we consider a special case that a $L=1$, $S=1$, $J=0$ dark matter pair annihilates into the $L=0$, $S=0$, $J=0$ final states (e.g., $\chi \chi \rightarrow \Phi \rightarrow f\overline{f}$, where $\chi$ is the majorana fermionic dark matter, $\Phi$ is an s-channel CP-even scalar mediator, and $f$ is the final product). In the section \ref{WaveFunction}, we will prove that the $L=1$, $S=1$, $J=0$ final states can also be described by the same equation. In this case, the index of the derivative of a $\delta$-function is contracted with the triplet spin-index of the initial state. Extract the third component of the initial-state triplet wave functions, we acquire
\begin{eqnarray}
-\frac{\vec{\nabla}^2}{2 \mu} \chi(\vec{r})+V(\vec{r}) \chi(\vec{r}) + u_p   \frac{\partial \delta^{(3)} (\vec{r})}{\partial z}  \phi(\vec{0}) &=& E_k \chi(\vec{r}) \nonumber \\
-\vec{\nabla}^2 \phi(\vec{r}) + \mu u_p \delta^{(3)}(\vec{r}) \left. \frac{\partial \chi(\vec{r}^{\prime})}{\partial z} \right|_{\vec{r}^{\prime}=\vec{0}} &=& \frac{E_t^2}{4} \phi(\vec{r}). \label{pFunction}
\end{eqnarray}
In this paper, we only solve (\ref{pFunction}). One can consider other cases. From the operators described in Ref.~\cite{pWaveOperator} (many of the similar operators also appear in the Ref.~\cite{pWaveOperatorQCD}), and follow the similar steps in our section \ref{WaveFunction}, we show the equations of other angular momentum combinations.  For $L=1$, $S=0$, $J=1$ dark matter annihilating to the $L=1$, $S=0$, $J=1$ final states, we have
\begin{eqnarray}
-\frac{\vec{\nabla}^2}{2 \mu} \chi(\vec{r})+V(\vec{r}) \chi(\vec{r}) + u_p   \frac{\partial \delta^{(3)} (\vec{r})}{\partial x_i} \left. \frac{\phi(\vec{r}^{\prime})}{\partial x_i} \right|_{\vec{r}^{\prime}=\vec{0}} &=& E_k \chi(\vec{r}), \nonumber \\
-\vec{\nabla}^2 \phi(\vec{r}) + \mu u_p \frac{\partial \delta^{(3)}(\vec{r})}{\partial x_i} \left. \frac{\partial \chi(\vec{r}^{\prime})}{\partial x_i} \right|_{\vec{r}^{\prime}=\vec{0}} &=& \frac{E_t^2}{4} \phi(\vec{r}). \label{pFunction_L1S0J1}
\end{eqnarray}
For $L=1$, $S=1$, $J=1$ dark matter annihilating to the $L=1$, $S=1$, $J=1$ final states, we have
\begin{eqnarray}
-\frac{\vec{\nabla}^2}{2 \mu} \chi_i(\vec{r})+V(\vec{r}) \chi_i(\vec{r}) + u_p   \frac{\partial \delta^{(3)} (\vec{r})}{\partial x_j} \left. \frac{\phi_j(\vec{r}^{\prime})}{\partial x_i} \right|_{\vec{r}^{\prime}=\vec{0}} - u_p   \frac{\partial \delta^{(3)} (\vec{r})}{\partial x_j} \left. \frac{\phi_i(\vec{r})}{\partial x_j} \right|_{\vec{r}=\vec{0}}&=& E_k \chi_i(\vec{r}), \nonumber \\
-\vec{\nabla}^2 \phi_i(\vec{r}) + \mu u_p \frac{\partial \delta^{(3)}(\vec{r})}{\partial x_j} \left. \frac{\partial \chi_j(\vec{r}^{\prime})}{\partial x_i}  \right|_{\vec{r}^{\prime}=\vec{0}} -  \mu u_p \frac{\partial \delta^{(3)}(\vec{r})}{\partial x_j} \left. \frac{\partial \chi_i(\vec{r}^{\prime})}{\partial x_j} \right|_{\vec{r}^{\prime}=\vec{0}} &=& \frac{E_t^2}{4} \phi_i(\vec{r}), \label{pFunction_L1S1J1}
\end{eqnarray}
where here $\chi_i$ and $\phi_i$ take a vector index $i=1,2,3$. Einstein summation convention has also been also applied. $L=1$, $S=1$, $J=1$ can also annihilate to $L=1$, $S=0$, $J=1$ final states,
\begin{eqnarray}
-\frac{\vec{\nabla}^2}{2 \mu} \chi_i(\vec{r})+V(\vec{r}) \chi_i(\vec{r}) + u_p  \epsilon_{i j k} \frac{\partial \delta^{(3)} (\vec{r})}{\partial x_j} \left. \frac{\phi(\vec{r})}{\partial x_k} \right|_{\vec{r}=\vec{0}} &=& E_k \chi_i(\vec{r}), \nonumber \\
-\vec{\nabla}^2 \phi(\vec{r}) + \mu u_p \epsilon_{ijk} \frac{\partial \delta^{(3)}(\vec{r})}{\partial x_i} \left. \frac{\partial \chi_j(\vec{r}^{\prime})}{\partial x_k}  \right|_{\vec{r}^{\prime}} &=& \frac{E_t^2}{4} \phi_i(\vec{r}), \label{pFunction_L1S1J1_2}
\end{eqnarray}
where $\epsilon_{ijk}$ is the full anti-symmetric tensor. For the $L=1$, $S=1$, $J=2$ annihilating to $L=1$, $S=1$, $J=2$ final states,
\begin{eqnarray}
-\frac{\vec{\nabla}^2}{2 \mu} \chi_i(\vec{r})+V(\vec{r}) \chi_i(\vec{r}) + u_p  \frac{\partial \delta^{(3)} (\vec{r})}{\partial x_j} \left[ \frac{\phi_i(\vec{r})}{\partial x_j} +  \frac{\phi_j(\vec{r})}{\partial x_i} \right]_{\vec{r}=\vec{0}} &=& E_k \chi_i(\vec{r}), \nonumber \\
-\vec{\nabla}^2 \phi_i(\vec{r}) + \mu u_p \frac{\partial \delta^{(3)}(\vec{r})}{\partial x_j} \left[ \frac{\partial \chi_i(\vec{r}^{\prime})}{\partial x_j} + \frac{\chi_j(\vec{r}^{\prime})}{\partial x_i}  \right]_{\vec{r}^{\prime}=\vec{0}} &=& \frac{E_t^2}{4} \phi_i(\vec{r}). \label{pFunction_L1S1J2}
\end{eqnarray}
We should note that all the differences during the calculations can finally be absorbed into the renormalization parameter $k_{\Lambda 2}$, although we are not going to present the detailed derivation in this paper. In the above equations, $\mu$ is the reduced mass of the dark matter pair. $u_{s,p}$ are the coupling constants in the s-wave and p-wave scattering cases, respectively. Because of the conservation of the energy during the annihilation processes, the total energy $E_t = E_k + 4 \mu$, where $E_k$ is the total kinematic energy of the dark matter particle pair. $V(\vec{r})$ is the potential indicating the long-distance interaction between the dark matter particles. Here, we only concern the spherically symmetric potential, so that $V(\vec{r})=V(r)$. Usually $\lim\limits_{r \rightarrow \infty} V(r) = 0$ and $V(r)$ can have a pole with no more than order one at the origin.

Temporarily omitting the interaction terms, after separating the angular variables, the radial equations are given by
\begin{eqnarray}
-\frac{1}{2 \mu} \chi_{l_1}^{\prime \prime}(r) + \frac{l_1 (l_1+1)}{2 \mu r^2} \chi_{l_1}(r) + V(r) \chi_{l_1}(r) &=& E_k \chi_{l_1}(r), \label{ChiR} \\
-\phi_{l_2}^{\prime \prime}(r) + \frac{l_2 (l_2+1)}{r^2} \phi_{l_2}(r) &=& \frac{E_t^2}{4} \phi_{l_2}(r), \label{PhiR}
\end{eqnarray}
where $l_{1,2}$ indicate the quantum number of the total orbital angular momentum of the $\chi$ and $\phi$ partial waves respectively. The definitions of the $\chi_{l_1}$ and $\phi_{l_2}$ are given by
\begin{eqnarray}
\chi(\vec{r}) = \sum_{l_1 = 0}^{\infty} \sum_{m_1 = -l_1}^{l_1}  (2 l_1 + 1) i^l \frac{\chi_{l_1}(r)}{r} Y_{l m}(\theta, \phi), \nonumber \\
\phi(\vec{r}) = \sum_{l_2 = 0}^{\infty} \sum_{m_2 = -l_2}^{l_2}  (2 l_2 + 1) i^l \frac{\phi_{l_1}(r)}{r} Y_{l m}(\theta, \phi),	\label{PartialWave}
\end{eqnarray}
where $m$ is the orbital angular momentum quantum number along the z-direction, $Y_{l m}(\theta, \phi)$ are the spherical harmonics, and $\theta$, $\phi$ are the angles in the polar coordinate system. Note that in the (\ref{ChiR}-\ref{PartialWave}), the wave functions are for the bound-states. As for the scattering states, the $\frac{\chi_{l_1} (r)}{r}$ and $\frac{\phi_{l_1} (r)}{r}$ should be replaced with the $\frac{\chi_{l_1} (k_1, r)}{k_1 r}$ and $\frac{\phi_{l_1} (k_2, r)}{k_2 r}$, where $k_{1,2}$ are the corresponding momentum.

Now we consider the eqn. (\ref{ChiR}), and define $k^2 = 2 \mu E_k$, (\ref{ChiR}) becomes
\begin{eqnarray}
-\chi_{l_1}^{\prime \prime}(r) + l_1 (l_1+1) \chi_{l_1}(r) + 2 \mu V(r) \chi_{l_1}(r) &=& k^2 \chi_{l_1}(r). \label{ChiR_Changed}
\end{eqnarray}
For each $k \in \mathbb{R}$ and $k \geq 0$, there are two linearly independent solutions $u_k(r)$, $v_k(r)$, with the two different asymptotic boundary conditions 
\begin{eqnarray}
& u_{l_1}(k, r) \propto r^{l_1+1},~v_{l_1}(k, r) \propto \frac{1}{r^{l_1}}, & \text{ when } r \rightarrow 0. \nonumber \\
& u_{l_1}(k, r) \rightarrow \sin(k x + \delta_{l_1}) ,~v_{l_1}(k, r) \rightarrow -\cos(k x + \delta_{l_1}), & \text{ when } r \rightarrow \infty. \label{Asymptotic}
\end{eqnarray}

In order to calculate the cross section, we need to decompose the wave functions into the asymptotic incoming and outgoing wave functions $e^{\mp i (k x + \delta_{l_1}) }$. Generally, the incoming and outgoing wave functions of a particular momentum $k$ can be written in the form
\begin{eqnarray}
u_{l_1, \text{in}}(k, r) &=& (-v_{l_1}(k, r) - i u_{l_1}(k, r)) \nonumber \\ 
u_{l_1, \text{out}}(k, r) &=& (-v_{l_1}(k, r) + i  u_{l_1}(k, r)) \label{DefinitionInOut} 
\end{eqnarray}
If $V(r)$ is an attractive potential, there will be some further bound states with a scattered spectrum of $E_i = k_{(b)i}^2 < 0$, where $i$ is some index to distinct the different bound states. We still express these eigenfunctions as $u_{l_1}(k_i, r)$, and the normalization convention is defined as
\begin{eqnarray}
\int_0^{\infty} u_{l_1}(k_{(b)i}, r) u_{l_1}(k_{(b)j}, r) dr = \delta_{i j}.
\end{eqnarray}
The asymptotic behaviour in the $r \rightarrow 0 $ is the same as the $k^2 > 0$ cases, but the asymptotic behaviour in the $r \rightarrow \infty$ is calculated to be
\begin{eqnarray}
u_{l_1}(k_{(b)i}, r) \sim e^{ \sqrt{-k_{(b)i}^2} r}.
\end{eqnarray}

According to the Sturm–Liouville theory, the $\lbrace u_{l_1} (k, r) \rbrace$, including all of the $u_{l_1} (k_{(b)i}, r)$'s, form a set of complete orthogonal basis on the function space defined on the $r>0$ domain. We should note that the $v_{l_1}$ can also be expanded on the basis of $u_{l_1}$.

However, if we restore the above incoming and outgoing radial functions  to the three-dimensional wave function $\chi_{l_1}$ and $\phi_{l_1}$, there will be a divergence at the origin, making them difficult to be coupled with the interaction term in the (\ref{sFunction}) and (\ref{pFunction}). Since $\lbrace u_{l_1}(k, r) \rbrace$ is a set of complete orthogonal basis, and is regular at the origin, we can expand $u_{l_1, \text{in/out}}$ with the basis of $u_{l_1}$, and indirectly calculate the couplings at the origin with their aids. In order to calculate the inner product defined by the integration, consider two functions $u_1(r)$ and $u_2(r)$ with the eigenvalue $k$ and $k^{\prime}$, satisfying
\begin{eqnarray}
-u_1^{\prime \prime}(r) + \frac{l_1 (l_1+1)}{r^2} u_1(r) + 2 \mu V(r) u_1(r) &=& k^2 u_1(r), \label{U1} \\
-u_2^{\prime \prime}(r) + \frac{l_1 (l_1+1)}{r^2} u_2(r) + 2 \mu V(r) u_2(r) &=& k^{\prime 2} u_2(r), \label{U2}
\end{eqnarray}
(\ref{U1})$\times u_2$-(\ref{U2})$\times u_1$, and after a few steps, we acquire
\begin{eqnarray}
- \frac{ \left[ u_1^{\prime} (r) u_2 (r) - u_1(r) u_2^{\prime}(r)\right]_0^{\infty} }{k^2-k^{\prime 2}} = \int_{0}^{\infty} u_1(r) u_2(r) dr.
\end{eqnarray}
Now we are ready to calculate the $\int_{0}^{\infty} u_{l_1, \text{in}}(k, r) u_{l_1}(k^{\prime}, r) dr$ and $\int_{0}^{\infty} u_{l_1, \text{out}}(k, r) u_{l_1}(k^{\prime}, r) dr$. Notice that if we add a small imaginary part $\mp i \lambda$ ($\lambda>0$) to the $k$ when solving the (\ref{ChiR_Changed}), we will acquire the damping wave functions $u_{l_1, \text{in}}(k-i \lambda, r) \approx u_{l_1, \text{in}}(k, r) e^{-\lambda r}$ and $u_{l_1, \text{out}}(k+i \lambda, r) \approx u_{l_1, \text{out}}(k, r) e^{-\lambda r}$. Therefore, we acquire $\lim\limits_{r \rightarrow \infty} \left[ u^{\prime}_{l_1, \text{in, out}}(k\mp i \lambda, r) u_{l_1}(k^{\prime}, r) - u_{l_1, \text{in, out}}(k\mp i \lambda, r) u^{\prime}_{l_1}(k^{\prime}, r) \right] = 0$. Finally, the inner products are given by
\begin{eqnarray}
\int_{0}^{\infty} u_{l_1, \text{in}}(k, r) u_{l_1}(k^{\prime}, r) dr &=& \frac{ \left[ u^{\prime}_{l_1, \text{in}}(k, r) u_{l_1}(k^{\prime}, r) - u_{l_1, \text{in}}(k, r) u^{\prime}_{l_1}(k^{\prime}, r) \right]_{r \rightarrow 0} }{(k-i \lambda)^2 - k^{\prime 2}}, \nonumber \\
\int_{0}^{\infty} u_{l_1, \text{out}}(k, r) u_{l_1}(k^{\prime}, r) dr &=& \frac{ \left[ u^{\prime}_{l_1, \text{out}}(k, r) u_{l_1}(k^{\prime}, r) - u_{l_1, \text{out}}(k, r) u^{\prime}_{l_1}(k^{\prime}, r) \right]_{r \rightarrow 0} }{(k+i \lambda)^2 - k^{\prime 2}}.
\end{eqnarray}
If $k = k^{\prime}$, $u^{\prime}_{l_1, \text{in}}(k, r) u_{l_1}(k, r) - u_{l_1, \text{in}}(k, r) u^{\prime}_{l_1}(k, r)$ and $u^{\prime}_{l_1, \text{out}}(k, r) u_{l_1}(k, r) - u_{l_1, \text{out}}(k, r) u^{\prime}_{l_1}(k, r)$ become constants (these are called the ``Wronskian'') independent of $r$. From the asymptotic definition (\ref{Asymptotic}), we can acquire
\begin{eqnarray}
u_{l_1, \text{in, out}}(k, r) \approx \frac{-1}{A_{l_1, k} (k r)^{l_1}},~u_{l_1}(k, r) \approx \frac{A_{l_1, k} (k r)^{l_1+1}}{2 l_1+1}, \text{ as $r$ approaches 0}. \label{CorrectedAsymptoticInZero}
\end{eqnarray}
Here, $A_{l_1, k}$ is a real number according to our convention in (\ref{DefinitionInOut}). For simplicity, we can adjust the phases in the (\ref{Asymptotic}) to guarantee $A_{l_1, k}>0$. Then,
\begin{eqnarray}
\left[ u^{\prime}_{l_1, \text{in, out}}(k, r) u_{l_1}(k^{\prime}, r) - u_{l_1, \text{in, out}}(k, r) u^{\prime}_{l_1}(k^{\prime}, r) \right]_{r \rightarrow 0} = \frac{- k^{\prime {l_1}+1} A_{l_1, k^{\prime}}}{k^{l_1} A_{l_1, k}}.
\end{eqnarray}
Immediately, we acquire
\begin{eqnarray}
\int_{0}^{\infty} u_{l_1, \text{in}}(k, r) u_{l_1}(k^{\prime}, r) dr &=& \frac{ k^{\prime {l_1}+1} A_{l_1, k^{\prime}} }{ k^{l_1} A_{l_1, k} \left[  k^{\prime 2} - (k-i \lambda)^2 \right] }, \label{InnerProductResultIn} \\
\int_{0}^{\infty} u_{l_1, \text{out}}(k, r) u_{l_1}(k^{\prime}, r) dr &=& \frac{ k^{\prime {l_1}+1} A_{l_1, k^{\prime}} }{ k^{l_1} A_{l_1, k} \left[ k^{\prime 2} - (k+i \lambda)^2 \right]}. \label{InnerProductResultOut}
\end{eqnarray}
We can then calculate the normalization coefficient of the basis $\lbrace u_{l_1} (k, r) \rbrace$. Calculate $\frac{(\ref{InnerProductResultOut}) - (\ref{InnerProductResultIn})}{2 i}$, and take the limit $\lambda \rightarrow 0$, we can prove that
\begin{eqnarray}
\int_{0}^{\infty} u_{l_1}(k, r) u_{l_1}(k^{\prime}, r) dr = \frac{\pi}{2} \delta(k - k^{\prime}).
\end{eqnarray}
With this result, we can then expand $u_{l_1, \text{in, out}}$ on the basis $\lbrace u_{l_1} (k, r) \rbrace$. Considering both the contributions from the bound states and the continuous spectrum states, the result is
\begin{eqnarray}
u_{l_1, \text{in}}(k, r) &=& \sqrt{\frac{2}{\pi}} \int_0^{\infty} dk^{\prime} \frac{ k^{\prime {l_1}+1} A_{l_1, k^{\prime}} }{ k^{l_1}l A_{l_1, k} \left[  k^{\prime 2} - (k-i \lambda)^2 \right] } u_{l_1}(k^{\prime}, r) \nonumber \\
&+& \sum_{k_i}  \frac{ k_i^{ {l_1}+1} A_{l_1, k_i} }{ k^{l_1}l A_{l_1, k} \left[  k_i^{2} - (k-i \lambda)^2 \right] } u_{l_1}(k_i, r) , \label{ExpandResultIn_Pre} \\
u_{l_1, \text{out}}(k, r) &=& \sqrt{\frac{2}{\pi}} \int_0^{\infty} dk^{\prime} \frac{ k^{\prime {l_1}+1} A_{l_1, k^{\prime}} }{ k^{l_1} A_{l_1, k} \left[  k^{\prime 2} - (k+i \lambda)^2 \right] } u_{l_1}(k^{\prime}, r) \nonumber \\
&+& \sum_{k_i}  \frac{ k_i^{ {l_1}+1} A_{l_1, k_i} }{ k^{l_1}l A_{l_1, k} \left[  k_i^{2} - (k+i \lambda)^2 \right] } u_{l_1}(k_i, r). \label{ExpandResultOut_Pre}
\end{eqnarray}
For abbreviation, it is convenient to define
\begin{eqnarray}
\sqrt{\frac{2}{\pi}} \int_{k^{\prime}} dk^{\prime} = \sqrt{\frac{2}{\pi}} \int_{0}^{\infty} dk^{\prime} + \sum_{k_i}, 
\end{eqnarray}
then
\begin{eqnarray}
u_{l_1, \text{in}}(k, r) &=& \sqrt{\frac{2}{\pi}} \int_{k^{\prime}} dk^{\prime} \frac{ k^{\prime {l_1}+1} A_{l_1, k^{\prime}} }{ k^{l_1}l A_{l_1, k} \left[  k^{\prime 2} - (k-i \lambda)^2 \right] } u_{l_1}(k^{\prime}, r), \label{ExpandResultIn} \\
u_{l_1, \text{out}}(k, r) &=& \sqrt{\frac{2}{\pi}} \int_{k^{\prime}} dk^{\prime} \frac{ k^{\prime {l_1}+1} A_{l_1, k^{\prime}} }{ k^{l_1} A_{l_1, k} \left[  k^{\prime 2} - (k+i \lambda)^2 \right] } u_{l_1}(k^{\prime}, r). \label{ExpandResultOut}
\end{eqnarray}
The $\phi_{l_2}$'s have the similar results, and the above discussions can be directly transferred to the $\phi_{l_2}$.

\section{S-Wave Scattering} \label{SWaveCalc}

\subsection{No Sommerfeld Effect} \label{NoSommerfeld_sWave}

In order for the comparison and preparation, we at first calculate the S-wave scattering process without Sommerfeld effect, i.e., $V(r)=0$, and $l_1=l_2=0$. In this case,
\begin{eqnarray}
& &u_0(k, r)=\sin(k r), ~~v_0(k,r)=\cos(k r); \nonumber \\
& &u_{0, \text{in}}(k, r)=e^{-i k r}, ~~v_0(k, r)=e^{i k r}.
\end{eqnarray}
The Fourier expansion then becomes
\begin{eqnarray}
u_{0, \text{in}}(k, r) &=& \sqrt{\frac{2}{\pi}} \int_{0}^{\infty} dk^{\prime} \frac{k^{\prime}}{k^{\prime 2} - (k-i \lambda)^2} u_0(k^{\prime}, r), \nonumber \\
u_{0, \text{out}}(k, r) &=& \sqrt{\frac{2}{\pi}} \int_{0}^{\infty} dk^{\prime} \frac{k^{\prime}}{k^{\prime 2} - (k+i \lambda)^2} u_0(k^{\prime}, r).
\end{eqnarray}
Just like in the quantum mechanics case, we write the $\chi_0$ and the $\phi_0$ in the composition of the incoming, reflecting and scattering wave functions
\begin{eqnarray}
\chi_0(r) &=& u_{0, \text{in}}(k_1, r) + C u_{0, \text{in}}(k_1, r), \nonumber \\
\phi_0(r) &=& D u_{0, \text{out}}(k_2, r),
\end{eqnarray}
where
\begin{eqnarray}
k_2 = \mu + \frac{k_1^2}{8 \mu},
\end{eqnarray}
due to the conservation of the energy. Therefore,
\begin{eqnarray}
& & -\frac{1}{2 \mu} \chi_0^{\prime \prime}(r) + V(r) \chi_0(r) \nonumber \\
&=& \frac{1}{2 \mu}\sqrt{\frac{2}{\pi}} \int_{0}^{\infty} dk^{\prime} \left[ \frac{k^{\prime 3}}{k^{\prime 2} - (k_1-i \lambda)^2} u_0(k^{\prime}, r) + \frac{C k^{\prime 3}}{k^{\prime 2} - (k_1+i \lambda)^2} u_0(k^{\prime}, r) \right], \label{NoSommerfeldExpansionChi}\\
& & -\phi_0^{\prime \prime}(r) = \sqrt{\frac{2}{\pi}} \int_{0}^{\infty} dk^{\prime}   \frac{D k^{\prime 3}}{k^{\prime 2} - (k_2+i \lambda)^2} u_0(k^{\prime}, r).  \label{NoSommerfeldExpansionPhi}
\end{eqnarray}
To solve the (\ref{sFunction}), we also need to expand the three-dimensional $\delta$-function,
\begin{eqnarray}
\delta^{(3)}(\vec{r}) = \int_0^{\infty} \frac{1}{2 \pi^2} \frac{\sin(k r)}{r} k dk, \label{DeltaExpasionNoSommerfeld}
\end{eqnarray}
and the zero-point value of the wave function is also calculated to be
\begin{eqnarray}
\chi(\vec{0}) &=& \sqrt{\frac{2}{\pi}} \int_{0}^{\infty} dk^{\prime} \left[ \frac{k^{\prime 2}}{k^{\prime 2} - (k_1-i \lambda)^2} + \frac{C k^{\prime 2}}{k^{\prime 2} - (k_1+i \lambda)^2} \right], \label{ZeroPointChiNoSommerfeld} \\
\phi(\vec{0}) &=& \sqrt{\frac{2}{\pi}} \int_{0}^{\infty} dk^{\prime}   \frac{D k^{\prime 2}}{k^{\prime 2} - (k_2+i \lambda)^2}, \label{ZeroPointPhiNoSommerfeld}
\end{eqnarray}
since $\lim\limits_{r \rightarrow 0} \frac{\sin(k r)}{r} = k$.

Note that the integral in the (\ref{ZeroPointChiNoSommerfeld}, \ref{ZeroPointPhiNoSommerfeld}) is divergent, some renormalization scheme is required. Let us calculate
\begin{eqnarray}
& & \int_{0}^{\infty} dk^{\prime} \frac{k^{\prime 2}}{k^{\prime 2} - (k \pm i \lambda)^2} = \int_{0}^{\infty} dk^{\prime} + \int_{0}^{\infty} \frac{k^2}{k^{\prime 2} - (k \pm i \lambda)^2} dk^{\prime} \nonumber \\
&=& \int_{0}^{\infty} dk^{\prime} + \frac{1}{2} \int_{-\infty}^{\infty} \frac{k^2}{k^{\prime 2} - (k \pm i \lambda)^2} dk^{\prime}. \label{ZeroPointNoSommerfeldRenormalization}
\end{eqnarray}
The last step is because the second integrand is an even function. The first divergent integral can be renormalized by a cut-off $k_{\Lambda}$, and the second integral is calculated to be $\mp \frac{i \pi k}{2}$. In this process, we have applied a different form of, but fundamentally equivalent renormalization scheme compared with the Ref.~\cite{Jackiw, RenormDelta1, Renormdelta2}. Therefore,
\begin{eqnarray}
\chi_{\text{r}}(\vec{0}) &=& \sqrt{\frac{2}{\pi}} \left[ (1+C) k_{\Lambda 1} + \frac{i \pi k_1}{2} (1-C) \right], \nonumber \\
\phi_{\text{r}}(\vec{0}) &=& \sqrt{\frac{2}{\pi}} \left[ D k_{\Lambda 2} - \frac{i \pi k_2}{2} D \right]. \label{ZeroPoint_Renormalized}
\end{eqnarray}
Then, with a few steps, we can see that if
\begin{eqnarray}
\frac{1+C}{2 \mu} + \frac{u_s}{2 \pi^2} D (k_{\Lambda 2} - \frac{i \pi k_2}{2}) = 0, \nonumber \\
D + \frac{u_s \mu}{2 \pi^2} \left[ (1+C) k_{\Lambda 1} + \frac{i \pi k_1}{2} (1-C) \right] = 0, \label{sWaveAlgebraEquation}
\end{eqnarray}
(\ref{sFunction}) can be satisfied for the eigenvalues $E_k = \frac{k_1^2}{2 \mu}$ and $E_{t}^2 = k_2^2$. Solving (\ref{sWaveAlgebraEquation}), we have
\begin{eqnarray}
C = \frac{-8 \pi^4 + (2 i k_{\Lambda 2} + k_2 \pi)(-2 i k_{\Lambda 1} + k_1 \pi ) u_s^2  \mu^2}{8 \pi^4 + (2 i k_{\Lambda 2} + k_2 \pi)(2 i k_{\Lambda 1} + k_1 \pi ) u_s^2  \mu^2}.
\end{eqnarray}
Notice that for a nonzero $u_s$, $C<1$, and for nonzero $k_{\Lambda 1,2}$, a phase shift will be induced, too. The cross sections for $\chi \rightarrow \chi$ scattering and $\chi \rightarrow \phi$ annihilation is
\begin{eqnarray}
\sigma_{\text{sc}} &=& \frac{4 \pi}{k_1^2} \left| \frac{C+1}{2 i} \right|^2, \label{sScatteringCrossSection}  \\
\sigma_{\text{ann}} &=& \frac{4 \pi}{k_1^2} \frac{1-|C|^2}{4}. \label{sAnnihilationCrossSection}
\end{eqnarray}
Although a matching with the quantum field theory perturbative calculations requires both the results from (\ref{sScatteringCrossSection}) and (\ref{sAnnihilationCrossSection}), we only write down the annihilation cross section explicitly because of our main topic,
\begin{eqnarray}
\sigma_{\text{ann}} = \frac{32 k_2 \pi^7 u_s^2 \mu^2}{k_1 [ 64 \pi^8 + 16 \pi^4(-4 k_{\Lambda_1} k_{\Lambda_2} + k_1 k_2 \pi^2) u_s^2 \mu^2 +  (4 k_{\Lambda 1}^2 + k_1^2 \pi^2)(4 k_{\Lambda 2}^2 + k_2^2 \pi^2) u_s^4 \mu^4 ] }. \label{sNoSommerfeldResult}
\end{eqnarray}

For the purpose of matching with the perturbative calculations of the quantum field theory, we also need to calculate the $\phi \rightarrow \phi$ cross sections. This is calculated to be
\begin{eqnarray}
C^{\prime} &=& -\frac{8 \pi^4 - 4 k_{\Lambda 1} k_{\Lambda 2} u_s^2 \mu^2 - 2 i k_2 k_{\Lambda 1} \pi u_s^2 \pi^2 + 2 i k_1 k_{\Lambda 2} \pi u_s^2 \mu^2 - k_1 k_2 \pi^2 u_s^2 \mu^2}{8 \pi^4 - 4 k_{\Lambda 1} k_{\Lambda 2} u_s^2 \mu^2 + 2 i k_2 k_{\Lambda 1} \pi u_s^2 \pi^2 + 2 i k_1 k_{\Lambda 2} \pi u_s^2 \mu^2 + k_1 k_2 \pi^2 u_s^2 \mu^2},\nonumber \\
\sigma_{\text{$\phi$sc}} &=& \frac{4 \pi}{k_1^2} \left| \frac{C^{\prime}+1}{2 i} \right|^2. \label{PhiSelfScatteringSWave}
\end{eqnarray}
For the brevity of this paper, we have skipped all of the lengthy but familiar derivation of the (\ref{PhiSelfScatteringSWave}).

\subsection{Sommerfeld Effect Calculations} \label{sResultsSommerfeldEffect}

If the Sommerfeld effect is considered, i.e., in the $V(r) \neq 0$ case, the $A_{0, k}$ should be calculated. Traditionally, a scattering wave function in the $V(r)$ potential $\psi(\vec{r})$ with the asymptotic condition $\psi(\vec{r}) \sim e^{i k z}$ $(r \rightarrow \infty)$ is calculated, and its origin value $|\psi_{k}(\vec{0})|$ is acquired for further calculations.  From the definitions in the (\ref{CorrectedAsymptoticInZero}), we learn that $A_{0, k}$ is actually the the $|\psi_{k}(\vec{0})|$. Therefore, the (\ref{NoSommerfeldExpansionChi}) should me modified to
\begin{eqnarray}
& & -\frac{1}{2 \mu} \chi_0^{\prime \prime}(r) + V(r) \chi_0(r) \nonumber \\
&=& \frac{1}{2 \mu}\sqrt{\frac{2}{\pi}} \int_{k^{\prime}} dk^{\prime} \left|\frac{\psi_{k^{\prime}}(\vec{0})}{\psi_{k}(\vec{0})} \right| \left[ \frac{k^{\prime 3}}{k^{\prime 2} - (k_1-i \lambda)^2} u_0(k^{\prime}, r) + \frac{C k^{\prime 3}}{k^{\prime 2} - (k_1+i \lambda)^2} u_0(k^{\prime}, r) \right]. \label{SommerfeldExpansionChi}
\end{eqnarray}
One might find out that the above equation is incorrect in the bound-state basis, since the $\psi_{\text{bound state}}(0)$ calculations are different with the scattering states $\psi_{k}(0)$. However, in this paper, for the brevity of the equations, we omit these differences, which does not disturb the results.  The $\delta$-function in the $\chi$-equation in the (\ref{sFunction}) can be written in the form
\begin{eqnarray}
\delta^{(3)}(\vec{r}) = \int_{k^{\prime}} |\psi_k(\vec{0})| \frac{1}{2 \pi^2} \frac{u_0(k, r)}{r} k dk, \label{DeltaExpasionSommerfeld}
\end{eqnarray}
(\ref{ZeroPointChiNoSommerfeld}) also changes the integral in (\ref{ZeroPointChiNoSommerfeld}) to be
\begin{eqnarray}
\chi(\vec{0}) &=& \sqrt{\frac{2}{\pi}} \int_{k^{\prime}} dk^{\prime} \left|\frac{\psi_{k^{\prime}}^2(\vec{0})}{\psi_{k_1}(\vec{0})} \right|  \left[ \frac{k^{\prime 2}}{k^{\prime 2} - (k_1-i \lambda)^2} + \frac{C k^{\prime 2}}{k^{\prime 2} - (k_1+i \lambda)^2} \right]. \label{ZeroPointChiSommerfeld} 
\end{eqnarray}
The integral can again be separated into two parts,
\begin{eqnarray}
& & \int_{k^{\prime}} dk^{\prime} \left|\frac{\psi_{k^{\prime}}^2(\vec{0})}{\psi_{k}(\vec{0})} \right| \frac{k^{\prime 2}}{k^{\prime 2} - (k \pm i \lambda)^2} \nonumber \\
&=&  \int_{k^{\prime}} \left|\frac{\psi_{k^{\prime}}^2(\vec{0})}{\psi_{k}(\vec{0})} \right| dk^{\prime} + \sum_{k_i} \left|\frac{\psi_{k_i}^2 (\vec{0})}{\psi_{k}(\vec{0})} \right| \frac{k^2}{k_i^{2} - (k \pm i \lambda)^2} dk^{\prime} \nonumber \\
&+& \frac{1}{2} \int_{-\infty}^{\infty} \left|\frac{\psi_{|k^{\prime}|}^2 (\vec{0})}{\psi_{|k|}(\vec{0})} \right| \frac{k^2}{k^{\prime 2} - (k \pm i \lambda)^2} dk^{\prime}. \label{ZeroPointSommerfeldRenormalization}
\end{eqnarray}
The (\ref{ZeroPointSommerfeldRenormalization}) has two problems. To integrate out the first term and compare it with the (\ref{ZeroPointNoSommerfeldRenormalization}), we need to know the $\psi_k(\vec{0})$ for every value of $k$, and the second term should not be calculated by simply picking up the residue only near the real axis because the full analytical properties of $\left|\frac{\psi_{k^{\prime}}(\vec{0})}{\psi_{k}(\vec{0})} \right|$ in the complex plane are unknown. However, the $k^{\prime} = \pm (k \pm i \lambda)$ residue contributions in the second term are proportional to $i (1-C)$, while the all the other parts are proportional to $(1+C)$ and can be attributed to one term $\frac{t}{\psi_k(\vec{0})} \cdot k_{\Lambda}$. $t$ can be calculated in any value of the cut off $k_{\Lambda}$ by principle. However, notice that if $k_{\Lambda} \gg k_1$, the integrand in the high-energy area mainly contribute to the results, and notice that because $\lim\limits_{k \rightarrow \infty} \psi_k(\vec{0}) = 1$, in this case $t \sim 1$. On the other hand, If $k_{\Lambda} \ll \mu$,  we will see the final result will be not sensitive to the detailed value of $t$. In the following discussions, we will still explicitly preserve $t$, and the (\ref{ZeroPoint_Renormalized}) can be modified to
\begin{eqnarray}
\chi_{\text{r}}(\vec{0}) &=& \sqrt{\frac{2}{\pi}} \left[ (1+C) \frac{t}{|\psi_{k_1}(\vec{0})|} k_{\Lambda 1} + |\psi_{k_1}(\vec{0})| \frac{i \pi k_1}{2} (1-C) \right], \nonumber \\
\phi_{\text{r}}(\vec{0}) &=& \sqrt{\frac{2}{\pi}} \left[ D k_{\Lambda 2} - \frac{i \pi k_2}{2} D \right].
\end{eqnarray}
After a similar calculation, the (\ref{sWaveAlgebraEquation}) now becomes
\begin{eqnarray}
& & \frac{1+C}{2 \mu |\psi_{k_1}(\vec{0})| } + \frac{u_s}{2 \pi^2} D (k_{\Lambda 2} - \frac{i \pi k_2}{2}) = 0, \nonumber \\
& & D + \frac{u_s \mu}{2 \pi^2} \left[ (1+C) \frac{t}{|\psi_{k_1}(\vec{0})|} k_{\Lambda 1} + |\psi_{k_1}(\vec{0})| \frac{i \pi k_1}{2} (1-C) \right] = 0.
\end{eqnarray}
Solving this equation, we have
\begin{eqnarray}
C = \frac{-8 \pi^4 + (2 i k_{\Lambda 2} + k_2 \pi)(k_1 \pi |\psi_{k_1}^2(\vec{0})| - 2 i k_{\Lambda 1} t) u_s^2 \mu^2}{8 \pi^4 + (2 i k_{\Lambda 2} + k_2 \pi)(k_1 \pi |\psi_{k_1}^2(\vec{0})| + 2 i k_{\Lambda 1} t) u_s^2 \mu^2},
\end{eqnarray}
It can be proved easily that $|C| \leq 1$ so the unitarity will never break up. and finally,
\begin{eqnarray}
& & \sigma_{\text{ann, Sommerfeld}}= \\
& & \frac{32 k_2 \pi^7 |\psi_{k_1} (\vec{0})|^2 u_s^2 \mu^2}{k_1 [ 64 \pi^8 + 16 \pi^4(-4 k_{\Lambda_1} k_{\Lambda_2} t + k_1 k_2 \pi^2 |\psi_{k_1} (\vec{0})|^2) u_s^2 \mu^2 + (4 k_{\Lambda 1}^2 t^2 + k_1^2 \pi^2 |\psi_{k_1} (\vec{0})|^4 )(4 k_{\Lambda 2}^2 + k_2^2 \pi^2) u_s^4 \mu^4 ] }. \label{SommerfeldFinalResult}
\end{eqnarray}
To observe the dependence of $\sigma_{\text{ann}}$ on the $|\psi_{k_1} (\vec{0})|^2$, we take the limit $k_{\Lambda 1}=k_{\Lambda 2}=0$, which is reasonable in some special cases to be described, then
\begin{eqnarray}
\sigma_{\text{ann, Sommerfeld}}=\frac{32 k_2 \pi^3 |\psi_{k_1} (\vec{0})|^2 u_s^2 \mu^2}{k_1 (8 \pi^2 + k_1 k_2 |\psi_{k_1} (\vec{0})|^2 u_s^2 \mu^2)^2}. \label{CutOffSetZero}
\end{eqnarray}
It is now clear that when $k_1 k_2 u_s^2 \mu^2 \ll 8 \pi^2$, $\sigma_{\text{ann, Sommerfeld}} \approx |\psi_{k_1} (\vec{0})|^2 \sigma_{\text{ann}}$ if $|\psi_{k_1} (\vec{0})|^2$ is not so large. For extremely large $|\psi_{k_1} (\vec{0})|^2$, $\sigma_{\text{ann, Sommerfeld}}$ then becomes saturated before reaching the unitarity bound.

\subsection{Some Discussions} \label{Discussions}

\begin{figure}
\includegraphics[width=1.5in]{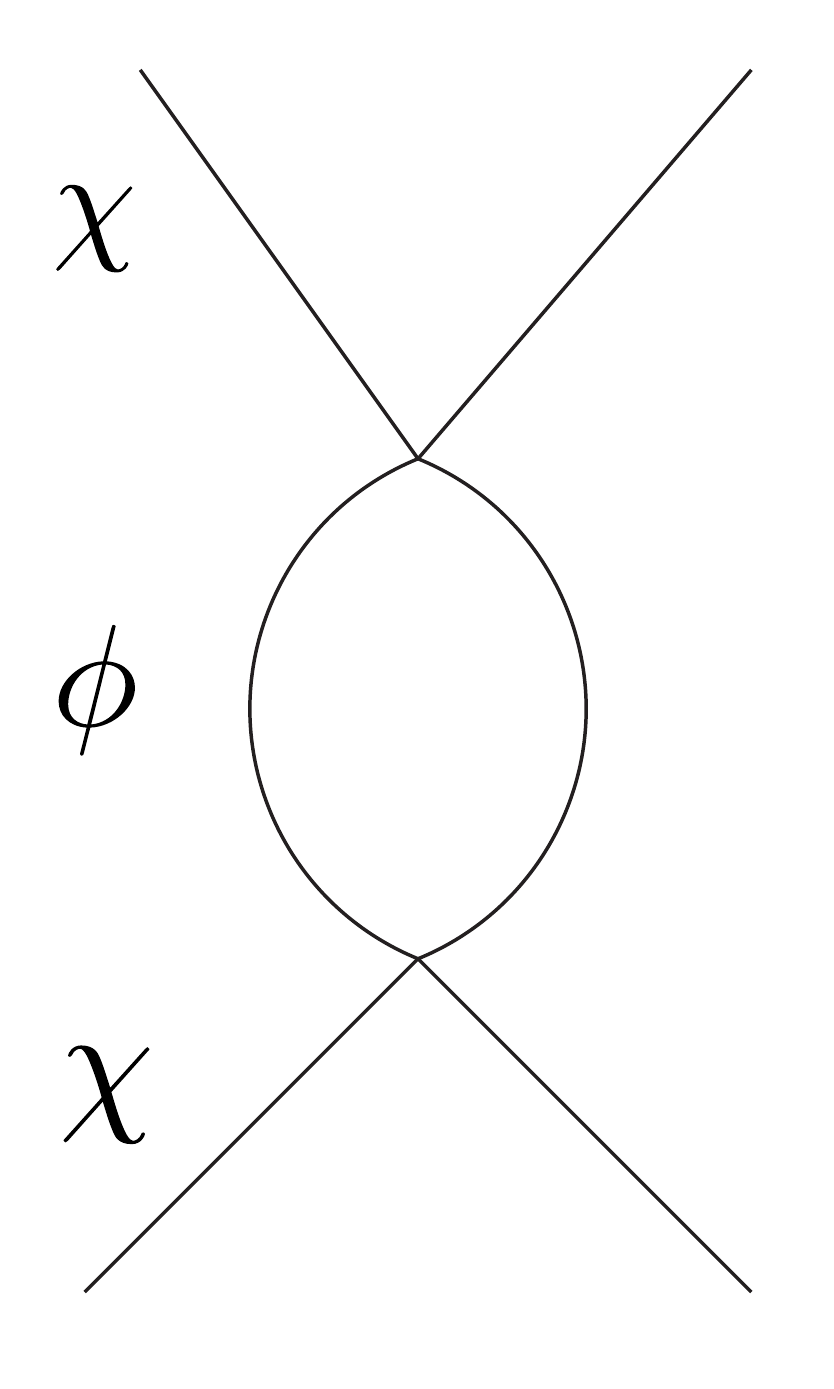}
\includegraphics[width=1.5in]{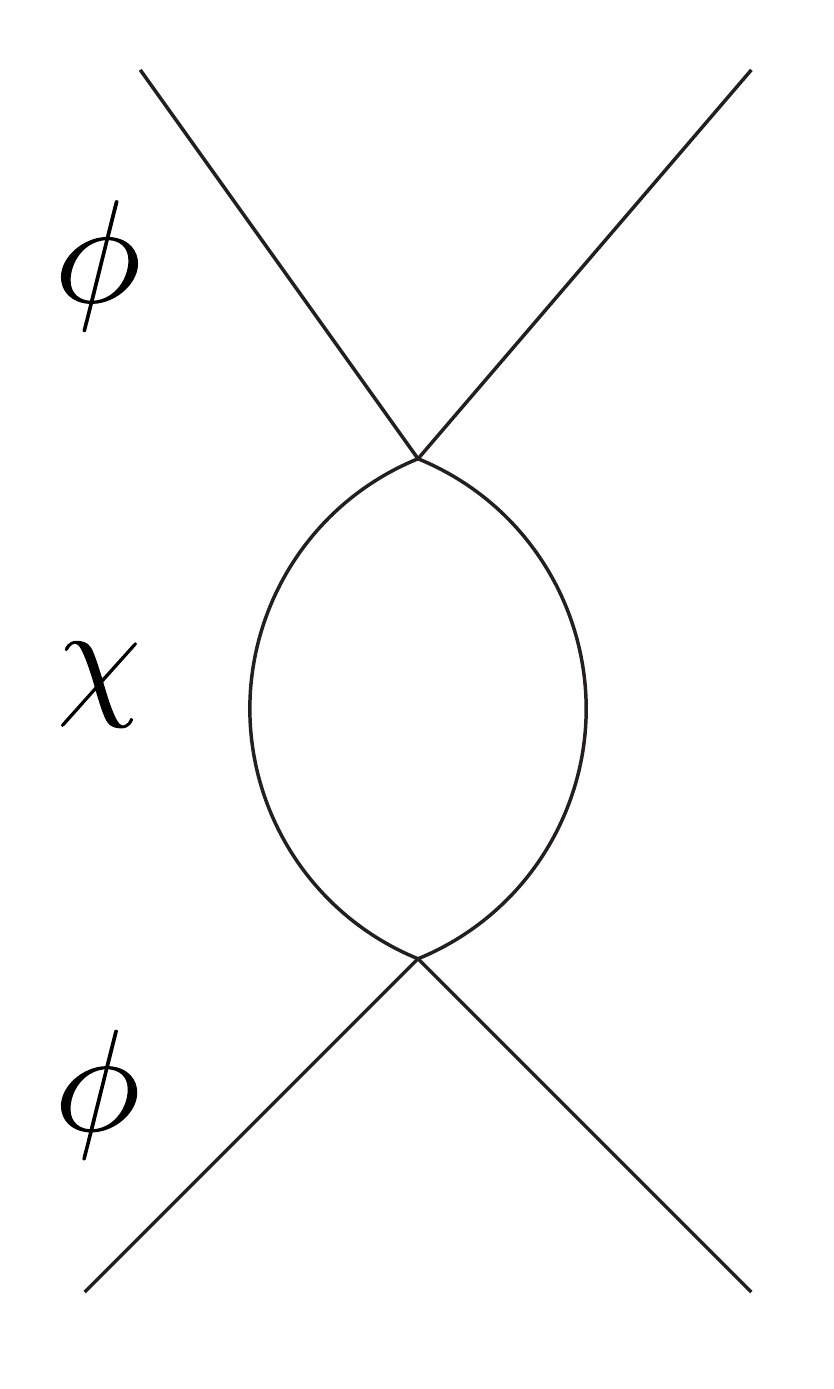}
\caption{Loop diagram that is considered when solving the Schrödinger equations. The time line is from the upward to downward.} \label{Divergences}
\end{figure}
\begin{figure}
\includegraphics[width=1.5in]{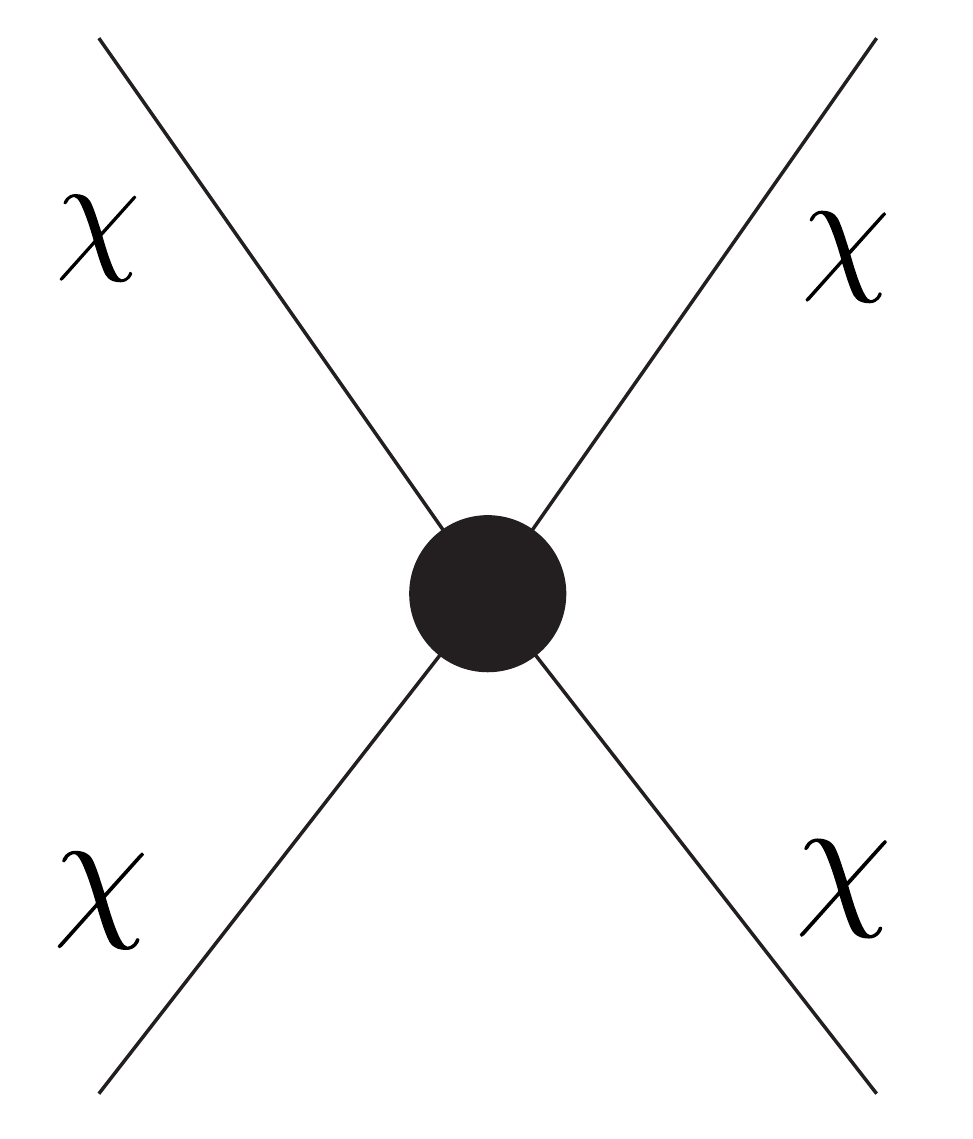}
\includegraphics[width=1.5in]{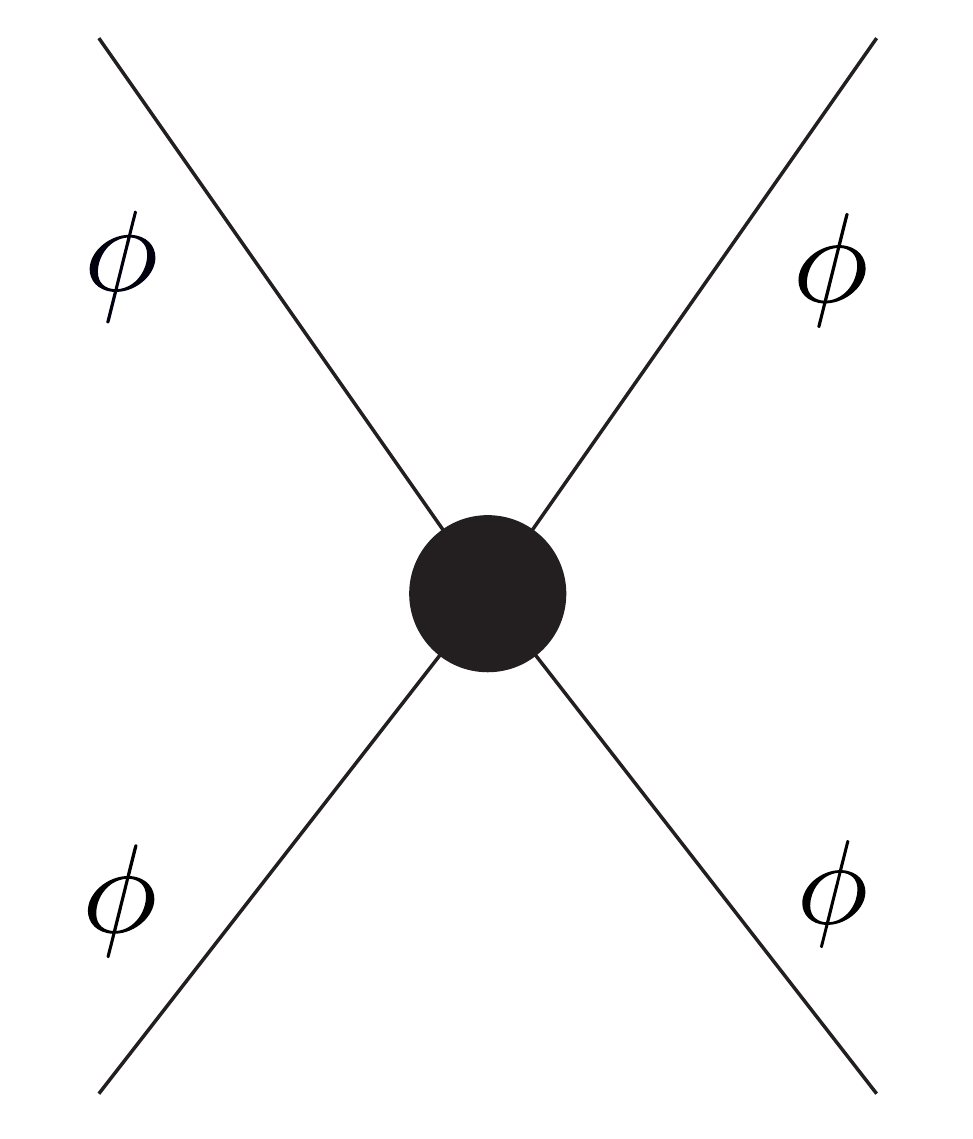}
\caption{Counter terms to cancel the divergences in the Fig.~\ref{Divergences}.} \label{CounterTerms}
\end{figure}

In this subsection, we will take some discussions on the renormalization scales. From the aspect of perturbative quantum field theory, the divergences arise from these two diagrams in the Fig.~\ref{Divergences}, which are corresponding to the $\chi(0)$ and $\phi(0)$ divergences. In the quantum field theory, the t- and u-channel loop diagrams are also considered. In the wave function calculations, the t- and u-channel loop diagrams are omitted, but the diagrams in the Fig.~\ref{Divergences} are connected from head to tail and are resummed. The divergences induced by these diagrams are cancelled out by the two counter-terms in the Fig.~\ref{CounterTerms}. One can try to add the corresponding self-interaction terms in the (\ref{sFunction}), however, this is equivalent to adjusting the renormalization scale $k_{\Lambda 1}$ and $k_{\Lambda 2}$.

It is easy to recognize that in the (\ref{sFunction}, \ref{pFunction}), we did not introduce the elastic short-distance terms indicating the $\chi \rightarrow \chi$ and $\phi \rightarrow \phi$ processes. However, such elastic scattering effects can still arise due to the Fig.~\ref{Divergences} loops together with the corresponding counter terms in the Fig.~\ref{CounterTerms}. Then we can calculate the Sommerfeld effect in a practical model by the following three steps:
\begin{itemize}
\item Calculate the short-distance perturbative s-wave dark matter annihilating cross section, dark matter s-wave self-interaction cross section, and the final state self-interaction cross section by the quantum field theory.
\item Compare the perturbative results with the (\ref{sScatteringCrossSection}, \ref{sNoSommerfeldResult}, \ref{PhiSelfScatteringSWave}) to determine the $u_s$, $k_{\Lambda 1}$, $k_{\Lambda 2}$ parameters.
\item Calculate the $\phi_{k_1}(\vec{0})$ as usual, and then determine the value of the $t$. Take all of these values to (\ref{SommerfeldFinalResult}) to calculate the Sommerfeld corrected cross section.
\end{itemize}

To determine the $t$, we need to know what we are comparing when we are discussing the ``boost factor''. Let us investigate the meaning of the cross section without the ``long-range'' interaction, i.e., the ``$\sigma v_0$'' from the perturbative quantum field theory aspect. Usually, if we shut down the long-range interactions to calculate the ``$\sigma v_0$'', all the divergences will change accordingly. Thus, we are actually comparing two models with totally different ``bare'' parameters.

\begin{figure}
\includegraphics[width=1.2in]{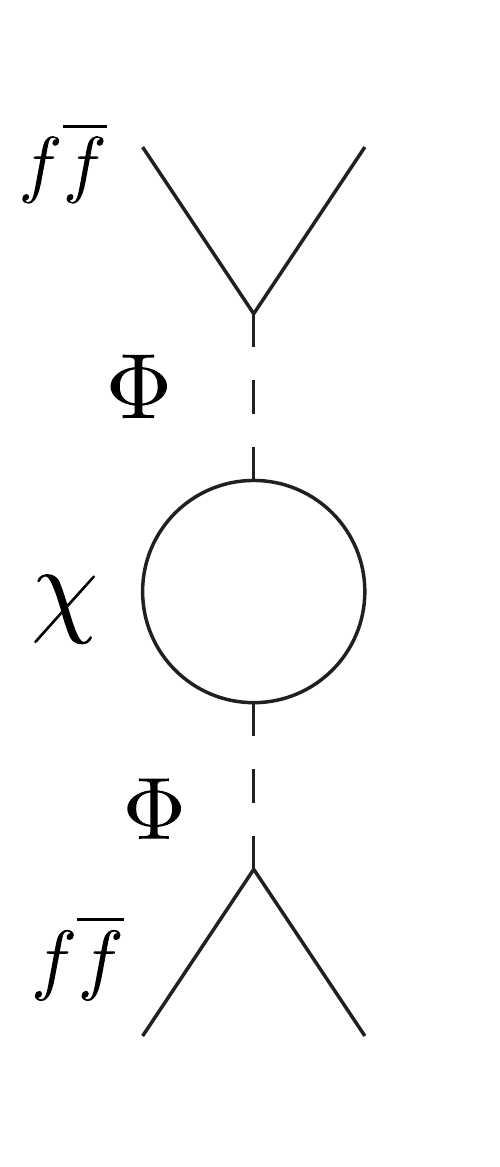}
\includegraphics[width=1.2in]{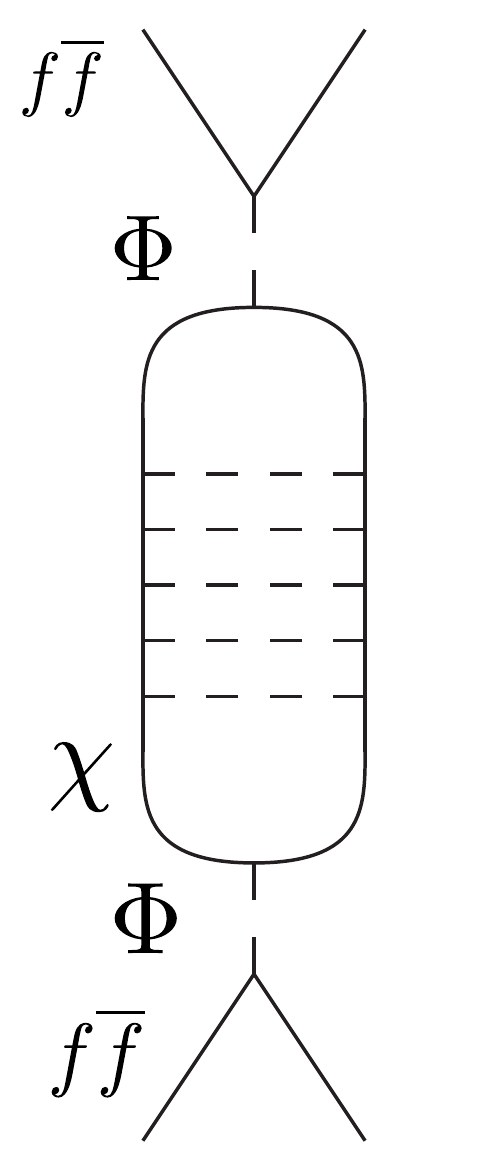}
\caption{The renormalization of a $\chi \chi \rightarrow \Phi \rightarrow f \overline{f}$ model.} \label{SpecialExample}
\end{figure}

For an example, let us consider a $\chi \chi \rightarrow \Phi \rightarrow f \overline{f}$ model, in which the dark matter $\chi$ is a majorana fermion, and $\Phi$ is an s-channel mediator, $f$ is the final product. In this model, the ``short-distance'' scattering is essentially described by a $\Phi$ propagator, and the divergences in the left panel of the Fig.~\ref{Divergences} are actually corresponding to the self-energy corrections to the $\Phi$, as shown in the Fig.~\ref{SpecialExample}. In a model with a ``long distance'' mediator, the appearance of the ladder loops at the right panel of the Fig.~\ref{SpecialExample} will substantially modify the divergence structures. It is natural to equalize the ``renormalized masses'' of the $\Phi$ mediator in both these models. Since the loop corrections on the mediator's mass are actually the divergent terms in the (\ref{NoSommerfeldExpansionChi}), (\ref{NoSommerfeldExpansionPhi}) and (\ref{SommerfeldExpansionChi}), considering the counter terms, and omitting some coupling constant factors, the corrections on the mediator's masses are finally the $k_{\Lambda 1}$ and $t k_{\Lambda 1}$ in the Sommerfeld or No-Sommerfeld cases respectively. Therefore, in order for a equivalent renormalized mass, we immediately know that $t=1$.

Such an analysis faces difficulty in some other models, e.g., the t-channel annihilation models, since the ``physical meanings'' of the divergences are not so obvious in this case. However, another approach to determine the $t$ can be found in the Ref.~\cite{MainDependence}. The authors of the Ref.~\cite{MainDependence} define the Sommerfeld factor by comparing the low-energy and the high-energy cross sections of exactly the same model. This prevent the problem of comparing two different models, and is actually equivalent to the $t=1$. Since this criteria is independent of detailed dark matter models, in this paper, we recommend that $t=1$.

If the counter terms ``coincidently'' cancels nearly all the divergence terms, the elastic scattering can be nearly prohibited. In this case, $k_{\Lambda 1,2} \ll \mu$. This can be an approximation for the models such like the dark matter annihilates into some light mediators through t-channel diagrams. In this case, the dark matter's self-interactions are mainly due to the long-distance effects. Therefore, our discussions in the (\ref{CutOffSetZero}) are available in such cases.

In this subsection, we also note that our renormalization process is completely different from the Ref.~\cite{MainDependence}. In the Ref.~\cite{MainDependence}, the zero-point wave functions (so-called the $G(0)$) is calculated by directly the $\frac{g_p (r)}{r}$ in the $r \rightarrow 0$ limit. This will result a $\log \frac{p}{p_0}$ term in their Eqn.~(32), which is referred to be unphysical in the Ref.~\cite{ZREFT}. Compared with their methods of a direct calculation near the singularity, we Fourier decomposite the ``unusual'' wave functions $v_{l_1}$ with the basis $u_{l_1}$, which are analytical in the $r=0$ area, thus we have avoided the singularity. What is more, the zero-point wave functions are actually equivalent to the Green's function $\langle 0 | T \psi(x_1) \psi(x_1) \psi(x_2) \psi(x_2) | 0 \rangle$, and our Eqn.~(\ref{ZeroPointSommerfeldRenormalization}) is compatible with the Eqn.~(2.39) in the Ref.~\cite{PetrakiBoundState1}( Notice that $\int d^3 q = \int_q 4 \pi q^2 dq$ in the  Ref.~\cite{PetrakiBoundState1}.) Therefore, we believe that our calculations are reasonable.

\section{P-Wave Scattering} \label{PWaveCalc}

Now, we are going to calculate the stationary scattering states of the (\ref{pFunction}). Since the derivative of a three-dimensional $\delta$-function at the origin only has a $l=1$ expansion along the $\theta-\phi$ direction, and particularly, $\frac{\partial \delta^{(3)}(\vec{r})}{\partial_z}$ only has a $l=1$, $m=0$ expansion, (\ref{pFunction}) can only couple the p-wave in the $\chi$-field with the s-wave $\phi$-field. Therefore, $l_1 = 1$ and $l_2 = 0$ in this case.

\subsection{No Sommerfeld Effect} \label{pResultNoSommerfeldResults}

Again, let us try $V(r)=0$ at first. In this case,
\begin{eqnarray}
u_1(k, r) =& k r j_1 (k r) &= \frac{\sin (k r)}{k r} - \cos (k r), \nonumber \\
u_2(k,r) =& k r j_{-2} (k r) &= -\frac{\cos (k r)}{k r} - \sin (k r),
\end{eqnarray}
where $j_n(x) = (-x)^n \left( \frac{1}{x} \frac{d}{dx} \right)^n \frac{\sin x}{x}$ are the spherical Bessel functions. The generalized Fourier expansion of the incoming and outgoing wave functions are given by
\begin{eqnarray}
u_{1, \text{in}}(k, r) &=& \sqrt{\frac{2}{\pi}} \int_{0}^{\infty} dk^{\prime} \frac{k^{\prime 2}}{k \left[ k^{\prime 2} - (k-i \lambda)^2 \right]} u_0(k^{\prime}, r), \nonumber \\
u_{1, \text{out}}(k, r) &=& \sqrt{\frac{2}{\pi}} \int_{0}^{\infty} dk^{\prime} \frac{k^{\prime 2}}{k \left[ k^{\prime 2} - (k+i \lambda)^2 \right]} u_0(k^{\prime}, r).
\end{eqnarray}
Define
\begin{eqnarray}
\chi_1(r) &=& u_{1, \text{in}}(k_1, r) + C u_{1, \text{in}}(k_1, r), \nonumber \\
\phi_1(r) &=& D u_{1, \text{out}}(k_2, r),
\end{eqnarray}
and then,
\begin{eqnarray}
& & -\frac{1}{2 \mu} \chi_1^{\prime \prime}(r) + \frac{2}{2 \mu r^2} \chi_1(r) +  V(r) \chi_1(r) \nonumber \\
&=& \frac{1}{2 \mu}\sqrt{\frac{2}{\pi}} \int_{0}^{\infty} dk^{\prime} \left[ \frac{k^{\prime 4}}{k \left[k^{\prime 2} - (k_1-i \lambda)^2\right]} u_0(k^{\prime}, r) + \frac{C k^{\prime 4}}{k \left[ k^{\prime 2} - (k_1+i \lambda)^2 \right]} u_0(k^{\prime}, r) \right]. \label{NoSommerfeldExpansionChi_pWave}
\end{eqnarray}
The derivative of the $\delta$-function is expanded to be
\begin{eqnarray}
\frac{\partial \delta^{(3)}(\vec{r})}{\partial z} = \partial_z \int_{0}^{\infty} \frac{1}{2 \pi^2} \frac{\sin(k r)}{r} k dk = \int_{0}^{\infty} \frac{1}{2 \pi^2}\frac{-k r j_1(k r)}{r} k^2 Y_{1 0}(\theta, \phi) dk.
\end{eqnarray}
The derivative of the wave function $\chi(\vec{r})$ at the origin is calculated to be
\begin{eqnarray}
\left. \frac{\partial \chi(\vec{r})}{\partial z} \right|_{\vec{r}=\vec{0}} &=& 2 i \sqrt{\frac{2}{\pi}} \int_{0}^{\infty} dk^{\prime} \left[ \frac{k^{\prime 4}}{k_1 \left[k^{\prime 2} - (k_1-i \lambda)^2\right]} + \frac{C k^{\prime 4}}{k_1 \left[ k^{\prime 2} - (k_1+i \lambda)^2 \right]} \right]. \label{ZeroPointChiNoSommerfeld_pWave} 
\end{eqnarray}
To separate the divergence,
\begin{eqnarray}
& & \int_{0}^{\infty} dk^{\prime} \frac{k^{\prime 4}}{k_1 \left[k^{\prime 2} - (k_1 \mp i \lambda)^2\right]} = \int_0^{\infty} dk^{\prime} \frac{k^{\prime 2}+k_1^2}{k_1} + \int_0^{\infty} dk^{\prime} \frac{k_1^4}{k_1 \left[ k_1^{\prime 2} - (k_1 \mp i \lambda)^2 \right] } \nonumber \\
&=& \int_0^{\infty} dk^{\prime} \frac{k^{\prime 2}+k_1^2}{k_1} \pm i k_1^2 \frac{i \pi}{2}.
\end{eqnarray}
The first term is renormalized to be $k_{\Lambda}^2$, so
\begin{eqnarray}
\left. \frac{\partial \chi_r(\vec{r})}{\partial z} \right|_{\vec{r}=\vec{0}} &=& 2 i \sqrt{\frac{2}{\pi}} \left[ (1+C) k_{\Lambda 1}^2 + \frac{i \pi k_1^2}{2} (1-C) \right].
\end{eqnarray}
The equations corresponding to the $\phi$ are similar to the s-wave cases, therefore we do not repeat them here.

Similar to the process in the section \ref{NoSommerfeld_sWave}, we can now write down the algebraic equations
\begin{eqnarray}
& &\frac{3 i}{2 \mu k_1} (1+C) - \frac{u_p}{2 \pi^2} (D k_{\Lambda 2} - D \frac{k_2 \pi i}{2}) = 0, \nonumber \\
& &D + \frac{u_p \mu}{2 \pi^2} 2 i \left[ (1+C) k_{\Lambda 1}^2 + \frac{i \pi k_1^2}{2} (1-C) \right] = 0.
\end{eqnarray}
The solution is
\begin{eqnarray}
C = \frac{-3 \pi^4 + k_1 (2 i k_{\Lambda 2} + k_2 \pi)(k_1^2 \pi - 2 i k_{\Lambda 1}^2 ) u_p^2 \mu^2}{3 \pi^4 + k_1 (2 i k_{\Lambda 2} + k_2 \pi)(k_1^2 \pi + 2 i k_{\Lambda 1}^2 ) u_p^2 \mu^2}.
\end{eqnarray}
In the p-wave case,
\begin{eqnarray}
\sigma_{\text{sc}} &=& \frac{36 \pi}{k_1^2} \left| \frac{C+1}{2 i} \right|^2, \label{pScatteringCrossSection}  \\
\sigma_{\text{ann}} &=& \frac{36 \pi}{k_1^2} \frac{1-|C|^2}{4}. \label{pAnnihilationCrossSection}
\end{eqnarray}
Correspondingly,
\begin{eqnarray}
\sigma_{\text{ann}} = \frac{9 \times 12 k_1 k_2 \pi^7 u_p^2 \mu^2}{9 \pi^8 + 6 k_1 \pi^4(-4 k_{\Lambda 1}^2 k_{\Lambda 2}+k_1^2 k_2 \pi^2) u_p^2 \mu^2 + k_1^2 (4 k_{\Lambda 1}^4 + k_1^4 \pi^2 )(4 k_{\Lambda 2}^2 + k_2^2 \pi^2) u_p^4 \mu^4}
\end{eqnarray}

Again, the $\phi \rightarrow \phi$ self scattering process can be calculated to be
\begin{eqnarray}
C^{\prime} &=& -\frac{12 k_1 \pi^4 + 4 k_{\Lambda 1}^2 k_{\Lambda 2} u_p^2 \mu^2 + 2 i k_2 k_{\Lambda 1}^2 \pi u_p^2 \mu^2 - 2 i k_1^2 k_{\Lambda 2} \pi u^2 \mu^2 + k_1^2 k_2 \pi^2 u_p^2 \mu^2}{12 k_1 \pi^4 + 4 k_{\Lambda 1}^2 k_{\Lambda 2} u_p^2 \mu^2 - 2 i k_2 k_{\Lambda 1}^2 \pi u_p^2 \mu^2 - 2 i k_1^2 k_{\Lambda 2} \pi u_p^2 \mu^2 - k_1^2 k_2 \pi^2 u_p^2 \mu^2}, \nonumber \\
\sigma_{\text{$\phi$sc}} &=& \frac{36 \pi}{k_1^2} \left| \frac{C^{\prime}+1}{2 i} \right|^2.
\end{eqnarray}

\subsection{Sommerfeld Effect Calculations} \label{pWaveResultsSommerfeldEffects}

When $V(r) \neq 0$, the p-wave Sommerfeld enhancement results depend on the $\frac{\partial \psi^{\prime}(\vec{0})}{\partial z}$. Since this has an extra imaginary unit and is a dimensional-1 parameter, we define $\psi^{\prime}_{k0} = \frac{1}{i k} \frac{\psi(\vec{0})}{\partial z}$, when $\psi$ is the stationary scattering wave function with the momenta $k$ along the z-direction. Again, from the (\ref{CorrectedAsymptoticInZero}), $A_{1,k} = |\psi^{\prime}_{k0}|$. The expansions of the incoming and outgoing wave functions have changed to
\begin{eqnarray}
& & -\frac{1}{2 \mu} \chi_1^{\prime \prime}(r) + \frac{2}{2 \mu r^2} \chi_1(r) +  V(r) \chi_1(r) \nonumber \\
&=& \frac{1}{2 \mu}\sqrt{\frac{2}{\pi}} \int_{k^{\prime}} dk^{\prime} \left|\frac{\psi_{k^{\prime} 0}^{\prime}}{\psi_{k 0}^{\prime}} \right|  \left[ \frac{k^{\prime 4}}{k \left[k^{\prime 2} - (k_1-i \lambda)^2\right]} u_1(k^{\prime}, r) + \frac{C k^{\prime 4}}{k \left[ k^{\prime 2} - (k_1+i \lambda)^2 \right]} u_1(k^{\prime}, r) \right]. \label{SommerfeldExpansionChi_pWave}
\end{eqnarray}
Here is the expansion of the $\frac{\partial \delta^{(3)} (\vec{r})}{\partial z}$,
\begin{eqnarray}
\frac{\partial \delta^{(3)}(\vec{r})}{\partial z} = \int_{k^{\prime}} \frac{|\psi_{k 0}^{\prime}|}{2 \pi^2}\frac{- u_1(k r)}{r} k^2 Y_{1 0}(\theta, \phi) dk,
\end{eqnarray}
and the derivative of the wave function $\chi(\vec{r})$ is modified to
\begin{eqnarray}
\left. \frac{\partial \chi(\vec{r})}{\partial z} \right|_{\vec{r}=\vec{0}} &=& 2 i \sqrt{\frac{2}{\pi}} \int_{k^{\prime}} dk^{\prime}  \left|\frac{\psi_{k^{\prime} 0}^{\prime 2}}{\psi_{k 0}^{\prime}} \right| \left[ \frac{k^{\prime 4}}{k_1 \left[k^{\prime 2} - (k_1-i \lambda)^2\right]} + \frac{C k^{\prime 4}}{k_1 \left[ k^{\prime 2} - (k_1+i \lambda)^2 \right]} \right]. \label{ZeroPointChiSommerfeld_pWave} 
\end{eqnarray}
After renormalizing the divergent part into $t k_{\Lambda 1}$, we have
\begin{eqnarray}
\left. \frac{\partial \chi_r(\vec{r})}{\partial z} \right|_{\vec{r}=\vec{0}} &=& 2 i \sqrt{\frac{2}{\pi}} \left[ (1+C) \frac{t}{|\psi_{k_1 0}^{\prime}|} k_{\Lambda 1}^2 + |\psi_{k_1 0}^{\prime}| \frac{i \pi k_1^2}{2} (1-C) \right].
\end{eqnarray}
Now, the algebraic equation is given by
\begin{eqnarray}
& &\frac{3 i}{2 \mu k_1 |\psi_{k_1 0}^{\prime}|} (1+C) - \frac{u_p}{2 \pi^2} (D k_{\Lambda 2} - D \frac{k_2 \pi i}{2}) = 0, \nonumber \\
& &D + \frac{u_p \mu}{2 \pi^2} 2 i \left[ \frac{t}{|\psi_{k_1 0}^{\prime}|} (1+C) k_{\Lambda 1}^2 + \frac{i \pi k_1^2}{2} |\psi_{k_1 0}^{\prime}| (1-C) \right] = 0.
\end{eqnarray}
The solution of this equation becomes
\begin{eqnarray}
C = \frac{-3 \pi^4 + k_1 (2 i k_{\Lambda 2} + k_2 \pi)(k_1^2 \pi |\psi_{k_1 0}^{\prime}|^2 - 2 i k_{\Lambda 1}^2 t ) u_p^2 \mu^2}{3 \pi^4 + k_1 (2 i k_{\Lambda 2} + k_2 \pi)(k_1^2 \pi |\psi_{k_1 0}^{\prime}|^2 + 2 i k_{\Lambda 1}^2 t ) u_p^2 \mu^2}. \label{C_In_PWave_Sommerfeld}
\end{eqnarray}
Correspondingly,
\begin{eqnarray}
& &\sigma_{\text{ann, Sommerfeld}} = \\
& & \frac{9 \times 12 k_1 k_2 \pi^7 |\psi_{k_1 0}^{\prime}|^2 u_p^2 \mu^2}{9 \pi^8 + 6 k_1 \pi^4(-4 k_{\Lambda 1}^2 k_{\Lambda 2} t +k_1^2 k_2 \pi^2 |\psi_{k_1 0}^{\prime}|^2 ) u_p^2 \mu^2 + k_1^2 (4 k_{\Lambda 1}^4 t^2 + k_1^4 \pi^2 |\psi_{k_1 0}^{\prime}|^4 )(4 k_{\Lambda 2}^2 + k_2^2 \pi^2) u_p^4 \mu^4}. \nonumber
\end{eqnarray}
The discussions of the renormalization is similar to the s-wave case, so we do not repeat it in this section. However, we should note that in a real dark matter model, loop diagrams in the Fig.~\ref{Divergences} will also contribute to the $|l=1, m=0\rangle \rightarrow |l=1, m=\pm 1\rangle$ amplitudes. This is done through the $\frac{\partial \delta^{(3)} (\vec{r})}{\partial z}$ terms which should by principle appear in the (\ref{pFunction}). The cross section of such a channel is at least proportional to $v^4$, where $v$ is the relative velocity of the two dark matter particles, so we do not concern this suppressed effect in this paper. Again, if the counter term cancels most of the divergences, we have
\begin{eqnarray}
& &\sigma_{\text{ann, Sommerfeld}} = \frac{9 \times 12 k_1 k_2 \pi^3 |\psi_{k_1 0}^{\prime}|^2 u_p^2 \mu^2}{(3 \pi^2 + k_1^3 k_2 |\psi_{k_1 0}^{\prime}|^2 u_p^2 \mu^2)^2}.
\end{eqnarray}
Again, it is obvious that the $C$ in the (\ref{C_In_PWave_Sommerfeld}) will be saturated before it reaches the unitarity bound.

\section{Wave Function of the Massless Final State particle Pair} \label{WaveFunction}

In this section, we are going to patch a leak. The first equations of the (\ref{sFunction}-\ref{pFunction}) are Schrödinger equations and there have been sufficient discussions about this in the literature (For an example and an application in the dark matter, see Ref.~\cite{Hisano2}). However, the final state particles can usually be considered as massless in most of the dark matter models, and whether the ``wave function of relative motion'' of such a relativistic system should satisfy the d’Alembert equation remains a quest. From the aspect of quantum field theory, the stationary state $|N\rangle$ is characterized by the Bethe-Salpeter wave function (For a review, see Ref.~\cite{BSWaveReview}. For the original paper, see Ref.~\cite{BS}. For an application in the dark matter bound state, see Ref.~\cite{PetrakiBoundState1}) defined by
\begin{eqnarray}
\psi_N(x_1, x_2, ..., x_n) = \langle N | T\left\lbrace \psi_1(x_1) \psi_2(x_2) ...\psi_n(x_n) \right\rbrace |0\rangle,
\end{eqnarray}
where $\langle N |$ is the corresponding n-particle stationary state, and $\psi_i(x_i)$ can be the same or different particle field operators. Generally, the time component $x_{1,2,...,n}^0$ can be different. However, if all the interactions among the particles can be regarded as instantaneous interactions, and the effective vertices are independent on the $q^0$, which is the time-like component of the transfer momentum, as just the case in our paper, the Bethe-Salpeter wave function can be reduced to the equal-time wave function
\begin{eqnarray}
\psi_N(t, \vec{r}_1, \vec{r}_2, ..., \vec{r}_n) = \langle N | \psi_1(t, \vec{r}_1) \psi_2(t, \vec{r}_2) ...\psi_n(r, \vec{r}_n) |0\rangle. \label{EqualTimeWaveFunction}
\end{eqnarray}
Solving these wave functions involve the Beth-Salpeter equation. However, in this paper we adopt another convenient method. We can directly take the derivation on $t$ at both sides of the (\ref{EqualTimeWaveFunction}), and then replace the operators on the right-handed side by the equation of motions.

As for the p-wave scattering processes into fermionic particle pairs, the final state can be either $S=0$ or $S=1$. The majorana dark matter particle pair's p-wave annihilation through an s-channel scalar particle usually result in $S=0$, or longitudinal $S=1$ final states. The following discussions are mainly based on this situation. Another possibility is that the complex scalar dark matter annihilating into transverse $S=1$ final states though an s-channel massive boson. We do not present the detailed derivation for brevity.

The massless fermionic final state particle pairs can be described by two Weyl spinors. Let $\psi_1(t, \vec{r})$ and $\psi_2(t, \vec{r})$ be the two 2-dimensional Weyl spinors, each of them satisfy the equation of motion
\begin{eqnarray}
\partial_t \psi_{1,2}(t, \vec{r}) =\vec{\nabla} \cdot \vec{\sigma} \psi_{1,2}(t, \vec{r}) + O(t, \vec{r}) (i \sigma^2 ) \psi_{2,1}^*(t, \vec{r}), \label{EquationMotionFermions}
\end{eqnarray}
where $\sigma^i$ are the Pauli matrices, and $O(\vec{r})$ is the interaction term containing the dark matter fields, e.g., for a s-wave interaction effective theory, $O(t, \vec{r})$ can be $g \Phi_1(t, \vec{r}) \Phi_2(t, \vec{r})$, where $\Phi(t, \vec{r})$ is the dark matter's field operator, $g$ is the coupling constant which can be complex, and for a p-wave interaction effective theory,  $O(t, \vec{r})=g \Phi_1(t, \vec{r}) \partial_i \Phi_2(t, \vec{r})$. The definition of the scalar-longitudinal-vector Bethe-Salpeter wave function is given by
\begin{eqnarray}
\psi_{w, 2}(t, \vec{r}_1, \vec{r}_2) = \langle N | \psi_1(t, \vec{r}_1) \psi_2^{T}(t, \vec{r}_2) (i \sigma^2)  |0\rangle, \label{DefinitionWaveFunctionFermions}
\end{eqnarray}
where $|N\rangle$ not only contains the $\psi_1$-$\psi_2$ state, the two-dark matter states $\Psi_1 \Psi_2$ are also included. Because we only care about the wave function of the relative motion of a particle pair, with the total momenta $\vec{p}_{\text{tot}}=\vec{0}$ at rest, we can learn that such a wave function only depends on the relative coordinate $\vec{r}_1-\vec{r}_2$, so we define
\begin{eqnarray}
\psi_{w, 2}(t, \vec{r}_1-\vec{r}_2) = \psi_{w, 2}(t, \vec{r}_1, \vec{r}_2) = \langle N | \psi_1(t, \vec{r}_1) \psi_2^{T}(t, \vec{r}_2) (i \sigma^2)  |0\rangle.
\end{eqnarray}
This wave function can be decomposed into a vector part and a scalar part
\begin{eqnarray}
\psi_{w, 2}(t, \vec{r}) = \vec{V}(t, \vec{r}) \cdot \vec{\sigma} + A(t, \vec{r}) I,
\end{eqnarray}
where $I$ is the $2 \times 2$ unitary matrix. Take the derivative the (\ref{DefinitionInOut}) according to the $t$ parameter and consider the (\ref{EquationMotionFermions}), we have
\begin{eqnarray}
& &\frac{\partial \psi_{w, 2}(t, \vec{r})}{\partial t} = \frac{\partial  \vec{V}(t, \vec{r})}{\partial t} \cdot \vec{\sigma} + \frac{\partial A(t, \vec{r})}{\partial t} I \nonumber \\
&=& 2 I (\vec{\sigma} \cdot \vec{\nabla}) A(t, \vec{r}) + (\vec{\sigma} \cdot \vec{\nabla} ) ( \vec{V} \cdot \vec{\sigma} ) + ( \vec{V} \cdot \vec{\sigma} ) (\vec{\sigma} \cdot \overleftarrow{\nabla} ) + \langle O(t, \vec{0}) \rangle D_w(\vec{r}), \nonumber \\
&=&2 I (\vec{\sigma} \cdot \vec{\nabla}) A(t, \vec{r}) +2  \vec{\nabla} \cdot \vec{V}(t, \vec{r}) + \langle O(t, \vec{0}) \rangle D_w (\vec{r}).
\label{EquationOfMotionBeforeSeparation}
\end{eqnarray}
Here, $ \langle O(t, \vec{0}) \rangle = \langle N | O(t, \vec{0}) | 0 \rangle$, and $D_w(\vec{r}) = \langle 0 | (i \sigma^2) \psi_2^*(t, \vec{r}) \psi_2^T(t, \vec{0}) (i \sigma^2) | 0 \rangle + \langle 0 | \psi_1(t, \vec{r}) \psi_1^{\dagger}(t, \vec{0})| 0 \rangle$. We should note that although we are calculating the (\ref{EquationOfMotionBeforeSeparation}) in the Heisenberg picture, we can still contract the operators of the same particle by the Wick's theorem since we have ignored all the $n>2$ multi-particle states in this paper.  Finally, $D_w$ is calculated to be
\begin{eqnarray}
D_w (\vec{r}) = \int \frac{d^4 p}{(2 \pi)^4} \left[ \frac{i (p^0 I + \vec{p} \cdot \vec{\sigma})}{p^2 + i \epsilon} e^{i \vec{p} \cdot \vec{r}} - \frac{i (i \sigma^2) (p^0 I + \vec{p} \cdot \vec{\sigma}^T) (i \sigma^2)}{p^2 + i \epsilon} e^{i \vec{p} \cdot \vec{r}} \right]= D^0_w(\vec{r}) I,
\end{eqnarray}
where
\begin{eqnarray}
D^0_w(\vec{r}) = \int \frac{d^4 p}{(2 \pi)^4} \frac{2 i p^0}{p^2 + i \epsilon} e^{i \vec{p} \cdot \vec{r}}.
\end{eqnarray}
The above derivation have applied the properties $(i \sigma^2) \sigma^{\mu T} (i \sigma^2) = \bar{\sigma}^{\mu}$.

Then we can separate the different $SO(3)$ representations in the (\ref{EquationOfMotionBeforeSeparation}), into spin-0 part $A$ and spin-1 part $\vec{V}$. We have
\begin{eqnarray}
\left\lbrace \begin{array}{l}
\frac{\partial  \vec{V}(t, \vec{r})}{\partial t} = 2 \vec{\nabla} A , \\
\frac{\partial A(t, \vec{r})}{\partial t} = 2 \vec{\nabla} \cdot \vec{V}(t, \vec{r}) + \langle O(t, \vec{0}) \rangle D_w^0 (\vec{r}).
\end{array}\right. \label{EquationInComponents}
\end{eqnarray}
Eliminating the $\vec{V}$, we acquire an equation about $A$,
\begin{eqnarray}
\frac{\partial^2 A(t, \vec{r})}{\partial t^2} = 4 \vec{\nabla}^2 A + \frac{\partial \langle O(t, \vec{0}) \rangle}{\partial t} D_{w}^0(\vec{r}).
\end{eqnarray}
Note that we are discussing a stationary state $\langle N |$. According to the Heisenberg's equation,
\begin{eqnarray}
\frac{\partial}{\partial t} \langle N | O(t, \vec{0} |0\rangle = \frac{1}{i} \langle N | [H, O(t, \vec{0}] |0\rangle = \frac{(E_N - E_0)}{i} \langle N | O(t, \vec{0} |0\rangle,
\end{eqnarray}
where $H$ is the Hamiltonian, and $E_N$, $E_0$ are the energy eigenvalue of the state $| N \rangle$ and $| 0 \rangle$ respectively. Usually, the vacuum energy $E_0$ is renormalized to be zero, so
\begin{eqnarray}
\frac{\partial^2 A(t, \vec{r})}{\partial t^2} = 4 \vec{\nabla}^2 A + \frac{E_N \langle O(t, \vec{0}) \rangle}{i} D_{w}^0(\vec{r}), \label{FermionResult}
\end{eqnarray}
which proved that the scalar part of the relative wave function of a massless fermionic particle pair does satisfy the d’Alembert equation with an interaction term.

As for the vector part $\vec{V}$, from the (\ref{EquationInComponents}), we can see that
\begin{eqnarray}
\frac{\partial \vec{\nabla} \times \vec{V}(t, \vec{r})}{\partial t} = 2 \vec{\nabla} \times \vec{\nabla} A = 0,
\end{eqnarray}
then
\begin{eqnarray}
\vec{\nabla} \times \vec{V}(t, \vec{r}) = \vec{f}(\vec{r}),
\end{eqnarray}
where $\vec{f}(\vec{r})$ is some vector field which is independent of $t$. Since we are only interested in stationary wave function with a positive energy $E$, there will be an unavoidable $e^{i E t}$ factor in the $\vec{V}(t, \vec{r})$. For a non-zero $\vec{V}(t, \vec{r})$, the only choice is that $\vec{f}(\vec{r}) = \vec{0}$. Therefore,
\begin{eqnarray}
\vec{\nabla} \times \vec{V}(t, \vec{r}) =\vec{0}.
\end{eqnarray}
This means that $\vec{V}$ should be a longitudinal vector field.

Notice that $A$ and $\vec{V}$ are not independent fields. Through (\ref{EquationInComponents}) we can see that they can be derived to each other. If $A$ is a s-wave ($L=0$) field, from (\ref{EquationInComponents}) we can see that $\vec{V}$ will indicate a $S=1$, $L=1$, $J=0$ case. Due to the similarity of the $S=0$ and the longitudinal $S=1$ field,  the $A$ and the longitudinal vector field $\vec{V}$ are actually equivalent in describing a $S=0$, $L=0$ or a longitudinal $S=1$, $L=1$, $J=0$ state. In fact, the interaction term in the (\ref{EquationMotionFermions}) can describe either the p-wave annihilations to the $S=0$, $L=0$ or to the longitudinal $S=1$, $L=1$, $J=0$ final states, depending on whether $g$ is real or imaginary. For simplicity, we can just choose the scalar $A$ as the final-state wave function in the (\ref{sFunction}, \ref{pFunction}).

In order to describe the p-wave annihilation to the transverse $S=1$ final states, we should construct the following Bethe-Salpeter wave function,
\begin{eqnarray}
\psi_{v, 11}(t, \vec{r}_1, \vec{r}_2) = \langle N | \psi_1 (t, \vec{r}) \psi_1^{\dagger} (t, \vec{r}) | 0 \rangle, \nonumber \\
 \psi_{v, 22}(t, \vec{r}_1, \vec{r}_2) = \langle N | \psi_2 (t, \vec{r}) \psi_2^{\dagger} (t, \vec{r}) | 0 \rangle. \label{TransverseWaveFunction}
\end{eqnarray}
These can again be decomposed into a scalar part $A^{\prime}_{11\text{, or } 22}(t, \vec{r})$ as well as a vector part $\vec{V}^{\prime}_{11\text{, or } 22}(t, \vec{r})$. However, some similar derivations can show us that in a stationary state with some non-zero frequency, the scalar part should be zero, and the vector part should satisfy the d'Alembert equation while keeping the transverse wave condition $\vec{\nabla} \cdot \vec{V}_{11\text{, or }22}(t, \vec{r}) = 0$. It is then obvious that (\ref{TransverseWaveFunction}) actually describes the transverse $S=1$ state. In this paper, we have omitted the detailed derivations of this case.

Now we are going to the Bosonic cases. For simplicity, we only discuss the real scalar field situation. Other situations should follow a similar process. Let $\phi_1 (t, \vec{r})$, $\phi_2 (t, \vec{r})$ be the two final state particle fields, and they satisfy the equation of motion
\begin{eqnarray}
\frac{\partial^2 \phi_{1,2}}{(\partial t)^2} = \vec{\nabla}^2 \phi_{1,2} + O(t, \vec{0}) \phi_{2,1}. \label{EquationMotionBosons}
\end{eqnarray}
The wave function is given by
\begin{eqnarray}
\psi_{s, 2}(t, \vec{r}_1, \vec{r}_2) = \langle N | \phi_1(t, \vec{r}_1) \phi_2(t, \vec{r}_2) |0\rangle. \label{DefinitionWaveFunctionBosons}
\end{eqnarray}
Again, we only discuss the wave function of relative motion so the wave function only depends on $\vec{r}_1-\vec{r}_2$,
\begin{eqnarray}
\psi_{s, 2}(t, \vec{r}_1-\vec{r}_2) = \psi_{s, 2}(t, \vec{r}_1, \vec{r}_2) = \langle N | \phi_1(t, \vec{r}_1) \phi_2(t, \vec{r}_2) |0\rangle. \label{RelativeWaveFunctionBosns}
\end{eqnarray}
Take the second order derivative on (\ref{RelativeWaveFunctionBosns}), we have
\begin{eqnarray}
& & \frac{\partial^2 \psi_{s, 2}(t, \vec{r}_1-\vec{r}_2)}{\partial t^2} \nonumber \\
&=& \langle N | \frac{\partial^2 \phi_1(t, \vec{r})}{\partial t^2} \phi_2(t, \vec{r}_2) |0\rangle + \langle N | \phi_2(t, \vec{r}_2)  \frac{\partial^2 \phi_2(t, \vec{r})}{\partial t^2} |0\rangle +  2 \langle N | \frac{\partial \phi_1(t, \vec{r})}{\partial t} \frac{\partial \phi_2(t, \vec{r}_2)}{\partial t} |0\rangle. \label{ScalarWaveFunctionUnresolved}
\end{eqnarray}
Directly applying the (\ref{EquationMotionBosons}) can solve the first two terms. The last cross-term is a little bit troublesome. To deal with this, notice that
\begin{eqnarray}
& & \frac{\partial}{\partial t} \left[ \langle N | \phi_1(t, \vec{r}_1) \frac{\partial \phi_2(t, \vec{r}_2)}{\partial t} |0\rangle - \langle N | \frac{\partial \phi_1(t, \vec{r}_1)}{\partial t} \phi_2(t, \vec{r}_2)  |0\rangle \right] \nonumber \\
&=& \langle N | \frac{1}{i} \left[ H, \phi_1(t, \vec{r}_1) \frac{\partial \phi_2(t, \vec{r}_2)}{\partial t}  -  \frac{\partial \phi_1(t, \vec{r}_1)}{\partial t} \phi_2(t, \vec{r}_2)  \right] |0\rangle \nonumber \\
&=& \frac{E_N}{i} \langle N | \phi_1(t, \vec{r}_1) \frac{\partial \phi_2(t, \vec{r}_2)}{\partial t}  -  \frac{\partial \phi_1(t, \vec{r}_1)}{\partial t} \phi_2(t, \vec{r}_2)  |0\rangle. \label{SinglePartial1}
\end{eqnarray}
On the other hand,
\begin{eqnarray}
& & \frac{\partial}{\partial t} \left[ \langle N | \phi_1(t, \vec{r}_1) \frac{\partial \phi_2(t, \vec{r}_2)}{\partial t} |0\rangle - \langle N | \frac{\partial \phi_1(t, \vec{r}_1)}{\partial t} \phi_2(t, \vec{r}_2)  |0\rangle \right] \nonumber \\
&=& \langle N | \phi_1(t, \vec{r}_1) \frac{\partial^2 \phi_2(t, \vec{r}_2)}{\partial t^2} |0\rangle - \langle N | \frac{\partial^2 \phi_1(t, \vec{r}_1)}{\partial t^2} \phi_2(t, \vec{r}_2)  |0\rangle \label{SinglePartial2}
\end{eqnarray}
Applying the equation of motion (\ref{EquationMotionBosons}) and notice that $\vec{\nabla}_1^2 = \vec{\nabla}_2^2$, the contraction $\langle 0 | \phi_1(t, \vec{r}_1)  \phi_1(t, \vec{r}_2) | 0 \rangle = \langle 0 | \phi_2(t, \vec{r}_1)  \phi_2(t, \vec{r}_2) | 0 \rangle$, then combining both (\ref{SinglePartial1}) and (\ref{SinglePartial2}), we have
\begin{eqnarray}
 \langle N | \phi_1(t, \vec{r}_1) \frac{\partial \phi_2(t, \vec{r}_2)}{\partial t} |0\rangle = \langle N | \frac{\partial \phi_1(t, \vec{r}_1)}{\partial t} \phi_2(t, \vec{r}_2) |0\rangle.
\end{eqnarray}
Therefore,
\begin{eqnarray}
 \langle N | \vec{\nabla}_1^2 \phi_1(t, \vec{r}_1) \frac{\partial \phi_2(t, \vec{r}_2)}{\partial t} |0\rangle &=& \langle N | \frac{\partial \vec{\nabla}_1^2 \phi_1(t, \vec{r}_1)}{\partial t} \phi_2(t, \vec{r}_2) |0\rangle, \nonumber \\
\langle N | \phi_1(t, \vec{r}_1) \frac{\partial \vec{\nabla}_2^2  \phi_2(t, \vec{r}_2)}{\partial t} |0\rangle &=& \langle N | \frac{\partial \phi_1(t, \vec{r}_1)}{\partial t} \vec{\nabla}_2^2 \phi_2(t, \vec{r}_2) |0\rangle.
\end{eqnarray}
As for the $\langle N| O(t, \vec{0}) \phi_1(t, \vec{x}_1) \frac{ \partial \phi_1(t, \vec{x}_2)}{\partial t} | 0 \rangle$ and the $\langle N| \frac{ \partial \phi_2(t, \vec{x}_2)}{\partial t}  O(t, \vec{0}) \phi_2(t, \vec{x}_1) | 0 \rangle$, since we are taking about some $\phi_1 \phi_2 |0\rangle$ states, and there should be no $\phi_1 \phi_1 |0\rangle$ state in the $|N\rangle$, the $\phi_1 \phi_1$ and $\phi_2 \phi_2$ terms can only contract to become $\langle 0 | \phi_1 \phi_1 |0 \rangle$ and $\langle 0 | \phi_2 \phi_2 |0 \rangle$. Notice that both $\phi_1$ and $\phi_2$ are massless real scalar particles, so the values of $\langle 0 | \phi_1 \phi_1 |0 \rangle$ and $\langle 0 | \phi_2 \phi_2 |0 \rangle$ should be equal. Therefore,
\begin{eqnarray}
& &\langle N| O(t, \vec{0}) \phi_1(t, \vec{x}_1) \frac{ \partial \phi_1(t, \vec{x}_2)}{\partial t} | 0 \rangle = \langle N | O(t, \vec{0}) | 0 \rangle \langle 0 | \phi_1(t, \vec{x}_1) \frac{ \partial \phi_1(t, \vec{x}_2)}{\partial t} | 0 \rangle \nonumber \\
&=& \langle N | O(t, \vec{0}) | 0 \rangle \langle 0 | \phi_2(t, \vec{x}_1) \frac{ \partial \phi_2(t, \vec{x}_2)}{\partial t} | 0 \rangle = \langle N| O(t, \vec{0}) \phi_2(t, \vec{x}_1) \frac{ \partial \phi_2(t, \vec{x}_2)}{\partial t} | 0 \rangle,
\end{eqnarray}
and then
\begin{eqnarray}
& & \frac{E}{i} \langle N| O(t, \vec{0}) \phi_1(t, \vec{x}_1) \phi_1(t, \vec{x}_2) | 0 \rangle  \nonumber \\
&=&  \frac{1}{i} \langle N| \left[ H, O(t, \vec{0}) \phi_1(t, \vec{x}_1) \phi_1(t, \vec{x}_2) \right] | 0 \rangle \nonumber \\
&=&  \frac{\partial}{\partial t} \langle N| O(t, \vec{0}) \phi_1(t, \vec{x}_1) \phi_1(t, \vec{x}_2) | 0 \rangle \nonumber \\
&=&  \langle N| \frac{\partial O(t, \vec{0})}{\partial t} |0 \rangle \langle 0| \phi_1(t, \vec{x}_1) \phi_1(t, \vec{x}_2) | 0 \rangle \nonumber \\
&+& \langle N|  O(t, \vec{0}) |0 \rangle \langle 0| \frac{\partial \phi_1(t, \vec{x}_1)}{\partial t} \phi_1(t, \vec{x}_2) | 0 \rangle  + \langle N|  O(t, \vec{0}) |0 \rangle \langle 0| \phi_1(t, \vec{x}_1) \frac{\partial \phi_1(t, \vec{x}_2)}{\partial t} | 0 \rangle \nonumber \\
&=& \frac{E}{i} \langle N| O(t, \vec{0}) \phi_1(t, \vec{x}_1) \phi_1(t, \vec{x}_2) | 0\rangle \nonumber \\
&+&  \langle N|  O(t, \vec{0}) |0 \rangle \langle 0| \frac{\partial \phi_1(t, \vec{x}_1)}{\partial t} \phi_1(t, \vec{x}_2) | 0 \rangle  + \langle N|  O(t, \vec{0}) |0 \rangle \langle 0| \phi_1(t, \vec{x}_1) \frac{\partial \phi_1(t, \vec{x}_2)}{\partial t} | 0 \rangle,
\end{eqnarray}
so
\begin{eqnarray}
0 &=&   \langle N|  O(t, \vec{0}) |0 \rangle \langle 0| \frac{\partial \phi_1(t, \vec{x}_1)}{\partial t} \phi_1(t, \vec{x}_2) | 0 \rangle  + \langle N|  O(t, \vec{0}) |0 \rangle \langle 0| \phi_1(t, \vec{x}_1) \frac{\partial \phi_1(t, \vec{x}_2)}{\partial t} | 0 \rangle \nonumber \\
&=& \langle N|  O(t, \vec{0})  \frac{\partial \phi_1(t, \vec{x}_1)}{\partial t} \phi_1(t, \vec{x}_2) | 0 \rangle  + \langle N|  O(t, \vec{0})  \phi_1(t, \vec{x}_1) \frac{\partial \phi_1(t, \vec{x}_2)}{\partial t} | 0 \rangle \nonumber \\
&=&\langle N|  O(t, \vec{0})  \frac{\partial \phi_1(t, \vec{x}_1)}{\partial t} \phi_1(t, \vec{x}_2) | 0 \rangle  + \langle N|  O(t, \vec{0})  \phi_2(t, \vec{x}_1) \frac{\partial \phi_2(t, \vec{x}_2)}{\partial t} | 0 \rangle.
\end{eqnarray}

Now, we are ready to calculate the cross term in the (\ref{ScalarWaveFunctionUnresolved}),
\begin{eqnarray}
& & \frac{E}{i} \langle N | \frac{\partial \phi_1(t, \vec{r})}{\partial t} \frac{\partial \phi_2(t, \vec{r}_2)}{\partial t} |0\rangle = \frac{1}{i} \langle N | \left[H, \frac{\partial \phi_1(t, \vec{r})}{\partial t} \frac{\partial \phi_2(t, \vec{r}_2)}{\partial t}\right] |0\rangle \nonumber \\
&=&\frac{\partial}{\partial t} \langle N | \frac{\partial \phi_1(t, \vec{r})}{\partial t} \frac{\partial \phi_2(t, \vec{r}_2)}{\partial t} |0\rangle \nonumber \\
&=& \langle N | \frac{\partial^2 \phi_1(t, \vec{r})}{\partial t^2} \frac{\partial \phi_2(t, \vec{r}_2)}{\partial t} |0\rangle + \langle N | \frac{\partial \phi_1(t, \vec{r})}{\partial t} \frac{\partial^2 \phi_2(t, \vec{r}_2)}{\partial t^2} |0\rangle \nonumber \\
&=& \langle N | \vec{\nabla}_1^2  \phi_1(t, \vec{r})\frac{\partial \phi_2(t, \vec{r}_2)}{\partial t} |0\rangle + \langle N | \vec{\nabla}_1^2 \frac{\partial \phi_1(t, \vec{r})}{\partial t}  \phi_2(t, \vec{r}_2) |0\rangle \nonumber \\
&+&\langle N|  O(t, \vec{0})  \frac{\partial \phi_1(t, \vec{x}_1)}{\partial t} \phi_1(t, \vec{x}_2) | 0 \rangle  + \langle N|  O(t, \vec{0})  \phi_2(t, \vec{x}_1) \frac{\partial \phi_2(t, \vec{x}_2)}{\partial t} | 0 \rangle \nonumber \\
&=& \frac{\partial}{\partial t} \langle N | \vec{\nabla}_1^2  \phi_1(t, \vec{r}) \phi_2(t, \vec{r}_2) |0\rangle = \frac{1}{i} \langle N | \left[ H, \vec{\nabla}_1^2  \phi_1(t, \vec{r}) \phi_2(t, \vec{r}_2) \right] |0\rangle \nonumber \\
&=& \frac{E}{i} \langle N | \vec{\nabla}_1^2  \phi_1(t, \vec{r}) \phi_2(t, \vec{r}_2) |0\rangle,
\end{eqnarray}
so
\begin{eqnarray}
\langle N | \frac{\partial \phi_1(t, \vec{r})}{\partial t} \frac{\partial \phi_2(t, \vec{r}_2)}{\partial t} |0\rangle = \langle N | \vec{\nabla}_1^2  \phi_1(t, \vec{r}) \phi_2(t, \vec{r}_2) |0\rangle. \label{CrossTermResults}
\end{eqnarray}
Combining (\ref{EquationMotionBosons}), (\ref{ScalarWaveFunctionUnresolved})and (\ref{CrossTermResults}), we have
\begin{eqnarray}
& & \frac{\partial^2 \psi_{s, 2}(t, \vec{r})}{\partial t^2} \nonumber \\
&=& 4 \vec{\nabla}^2 \psi_{s, 2}(t, \vec{r}) + 2 \langle O(t, \vec{r}) \rangle D_s(\vec{r}),
\end{eqnarray}
where $\langle O(t, \vec{r}) \rangle = \langle N | O(t, \vec{r}) | 0 \rangle$ and
\begin{eqnarray}
D_s(\vec{r}) = \langle 0 | \phi_1(t, \vec{r}), \phi_1(t, \vec{0}) 0 \rangle = \langle 0 | \phi_2(t, \vec{r}), \phi_2(t, \vec{0}) 0 \rangle = \int \frac{d^4 p}{(2 \pi)^4} \frac{i}{p^2+i \epsilon} e^{i \vec{p} \vec{r}}. \label{BosonicResult}
\end{eqnarray}
Here we have proved that the wave function of the massless scalar particle pair satisfies d’Alembert equation.

Finally, we need to explain the different orders of the divergences between the diagramatic and wave-function calculations. For example, if we calculate the loops in the  Fig.~\ref{Divergences}, we acquire a logarithmic divergence in the scalar case, and a quadratic result in the fermionic case. However, a wave-function based calculation gives a linear divergence. This is due to the following two reasons,
\begin{itemize}
\item In the wave-function approach, we describe the ``short-distance'' interactions by some ``point-like'' $\delta$-functions. This is usually not the truth, and will modify the orders of the divergences.
\item The ``ladder approximations'' of the wave-function approach usually pick up some of the poles of the particle propagators, while ignores some others. Although these ignored poles does not affect the nearly on-shell performances described by the wave functions, they will actually contribute to the divergence orders.
\end{itemize}

Fortunately, all these differences can be absorbed to the renormalization scale $k_{\Lambda 2}$, and since usually, the $k_2 \sim 2 \mu$ remains nearly unchanged when the dark matter relative velocity $v \ll 1$, we also do not need to worry about the running couplings which behaves quite differently in the bosonic and fermionic cases. Therefore, our calculations in this paper are still reasonable.

\section{Summary} \label{Summary}

In this paper, we have made the general discussions and calculations on the s-wave or the p-wave dark matter particle pair annihilating into a pair of s-wave massless particles. Practically, one can compare the results from the perturbative quantum field theory with the information of dark matter annihilation, dark matter elastic scattering and final state particle elastic scattering informations with the formulas in the section \ref{NoSommerfeld_sWave}, or \ref{pResultNoSommerfeldResults}, to determine the coupling constant $u$, and the two renormalization scales $k_{\Lambda 1,2}$, and then calculate the $\psi_k(\vec{0})$ and $\psi_{k0}^{\prime}$ by following the formulas in the section \ref{sResultsSommerfeldEffect}, or \ref{pWaveResultsSommerfeldEffects} for a complete calculation. We have also proved that the wave function of the final state can actually be described by the wave function equation (\ref{sFunction}) and (\ref{pFunction}).

\begin{acknowledgements}

Y.L.T. thanks for Pyungwon Ko, Peiwen Wu, Liping Sun, Chen Zhang, Dong Bai for helpful discussions. Y.L.T. is supported by the Korea Research Fellowship Program through the National Research Foundation of Korea (NRF) funded by the Ministry of Science and ICT (2017H1D3A1A01014127), and is also supported in part by National Research Foundation of Korea (NRF) Research Grant NRF-2015R1A2A1A05001869. G.L.Z. is supported by the Excellent Young Scientist Foundation of Xi'an University of Science and Technology under Grant No.~2018YQ3-15.

\end{acknowledgements}

\newpage
\bibliography{SommerfeldFinal}
\end{document}